\documentclass[compsoc, conference, a4paper, 10pt, times]{IEEEtran}
\usepackage{cite}
\usepackage{amsmath,amssymb,amsfonts}
\usepackage{graphicx}
\usepackage{textcomp}
\usepackage{xcolor}
\usepackage{booktabs}
\usepackage[hidelinks]{hyperref}

\usepackage{amsmath}

\usepackage{multirow}
\usepackage{adjustbox}
\usepackage{color}

\usepackage{booktabs}
\usepackage{multirow}
\usepackage{adjustbox}
\usepackage{listings}
\usepackage{xcolor}
\usepackage{framed}
\usepackage{subcaption}
\usepackage{comment}

\usepackage{array}
\usepackage{multirow}

\usepackage{algorithm}
\usepackage{algpseudocode}

\usepackage{subcaption}
\usepackage{wrapfig}

\usepackage{soul}

\begin{document}
\title{An Unbiased Transformer Source Code Learning with \\Semantic Vulnerability Graph}

\author{\em Anonymous Authors}











\author{\IEEEauthorblockN{Nafis Tanveer Islam\textsuperscript{1}, Gonzalo De La Torre Parra\textsuperscript{2}, Dylan Manuel\textsuperscript{1} \\Elias Bou-Harb\textsuperscript{1}, Peyman Najafirad\textsuperscript{1, *}\footnote{corresponding author}}

\IEEEauthorblockA{\textit{\textsuperscript{1}University of Texas at San Antonio, \textit{\textsuperscript{2}University of the Incarnate Word}} \\
\textsuperscript{1}nafistanveer.islam@utsa.edu, \textsuperscript{2}gdparra@uiwtx.edu, \textsuperscript{1}dylan.manuel@my.utsa.edu, \\\textsuperscript{1}elias.bouharb@utsa.edu, \textsuperscript{1}peyman.najafirad@utsa.edu}
\\

}


\maketitle

\newcommand\blfootnote[1]{%
  \begingroup
  \renewcommand\thefootnote{}\footnote{#1}%
  \addtocounter{footnote}{-1}%
  \endgroup
}

\begin{abstract}
Over the years, open-source software systems have become prey to threat actors. \blfootnote{* Corresponding Author} Even highly-adopted software has been crippled by unforeseeable attacks, leaving millions of devices exposed. Even as open-source communities act quickly to patch the breach, code vulnerability screening should be an integral part of agile software development from the beginning. Unfortunately, current vulnerability screening techniques are ineffective at identifying novel vulnerabilities or providing developers with code vulnerability and classification. \color{black} Furthermore, the datasets used for vulnerability learning often exhibit distribution shifts from the real-world testing distribution due to novel attack strategies deployed by adversaries and as a result, the machine learning model's performance may be hindered or biased\color{black}. To address these issues, we propose a joint interpolated multitasked unbiased vulnerability classifier comprising a transformer "RoBERTa" and graph convolution neural network (GCN). We present a training process utilizing a semantic vulnerability graph (SVG) representation from source code, created by integrating edges from a sequential flow, control flow, and data flow, as well as a novel flow dubbed Poacher Flow (PF). \color{black}Poacher flow edges reduce the gap between dynamic and static program analysis and handle complex long-range dependencies. Moreover, our approach reduces biases of classifiers regarding unbalanced datasets by integrating Focal Loss objective function along with SVG. \color{black} Remarkably, experimental results show that our classifier outperforms state-of-the-art results on vulnerability detection with fewer false negatives and false positives. After testing our model across multiple datasets, it shows an improvement of at least 2.41\% and 18.75\% in the best-case scenario. Evaluations \color{black} using N-day program samples demonstrate that our proposed approach achieves a 93\% accuracy and was able to detect 4, zero-day vulnerabilities from popular GitHub repositories. Our code and data are available at \url{https://github.com/pial08/SemVulDet} \color{black} 
\end{abstract}



\section{Introduction}
\label{1_introduction}

Threat actors exploit source code vulnerabilities using some of the same sophisticated techniques used in high-level cyberattack campaigns. In many cases, vulnerabilities are introduced by software developers using code snippets from open-source solutions, such as Stack Overflow and GitHub. With insufficient vulnerabilities, new software releases can be high risk. IBM \cite{IBM} findings suggest software vulnerabilities cost businesses an average of 3.9 million dollars annually; meanwhile, Common Vulnerabilities and Exposures (CVE) \cite{cve} reported that 6,015 new CVEs were added during the first quarter of 2022, a 36\% increase compared to 4,415 CVEs published in the first quarter of 2021. These reports demonstrate the ubiquity of these vulnerabilities and the importance of detecting and classifying them in large-scale applications.

Detecting vulnerabilities in large-scale programs at an early stage in software development is both a challenge and priority for software developers \cite{braz2022less}. Most software developers are experts at writing code, however, they are not always well-versed in software security \cite{gai2015proactive}, \cite{thakur2015investigation}. A recent zero-day exploit "Log4Shell" was announced against the Apache log4j library \cite{log4j}, the magnitude of which was unlike any other.  Given the large adoption of this Java logging library, threat actors had ripe opportunity to take control of web-facing servers, and even services not connected directly to the internet, by permeating malicious code to back-end software running Apache Log4j versions.
%
%
To prevent such cases, vulnerability classifiers can be enhanced to support developers, whether or not they are experts in security. Existing classifiers \cite{zhou2019devign, li2021sysevr, hou2022vulnerability} have been proposed for vulnerability detection in code snippets at an early stage of software development. However, these detectors only provide information on whether or not a vulnerability exists in a particular code snippet, but  no information is provided regarding the categories of their vulnerabilities.


\color{black}
The vulnerability distribution of real-world production software is imbalanced, as benign source code is released more frequently than vulnerable source code. According to Chakaraborty et al. \cite{chakraborty2021deep}, the suboptimal performance of current deep learning techniques in predicting real-world software vulnerabilities is attributed to training data imbalance and models. To address this issue in training datasets, several techniques such as data augmentation and syntactic data creation have been employed to mitigate the training data imbalance problem \cite{johnson2019survey}, \cite{chawla2002smote}. Although balanced training datasets through syntactic data generation are crucial, but they do not typically reflect the distribution shifts that are likely to cause from real-world. According to Geirhos, distribution shifts are underrepresented in the datasets widely used in the ML community today \cite{koh2021wilds}. This impacts the performance of the model in predicting real-world vulnerabilities such as N-day and zero-day.  
\color{black}

%

Recent graph creation techniques such as AST \cite{bilgin2020vulnerability} and Code Property Graph (CPG) \cite{zhou2019devign, yamaguchi2014modeling, wu2022code} are highly effective at detecting source code vulnerabilities. Even so, they have some bottlenecks. For example, neither AST nor CPG can capture the information when a variable is used out of its scope (when the variable is freed), since this is syntactically correct and only produced during the execution of a program. Similarly, when a divide-by-zero vulnerability occurs, it is exposed during program execution and is therefore also syntactically correct. In addition, current graph generation techniques \cite{zhou2019devign, bilgin2020vulnerability} fail to capture the long-range dependency of a variable. AST and CPG generate nodes and edges per statement, thus ignoring the long-range dependency between two faraway statements. Simply put, these graph-creation techniques fail to capture relatively simple yet high-frequency vulnerabilities, either because they occur during execution, or because of a lack of long-range dependency information.

Long-range dependency \cite{liu2019deepbalance} \cite{li2022automated} is a major challenge in vulnerability detection. A long-range dependency may be caused by a variable declared at the beginning of a function, but the vulnerability associated with that variable may only appear after a few hundred lines of code. Several transformer-based works were previously proposed to address the detection of vulnerabilities in source code \cite{ding2021velvet} \cite{hou2022vulnerability}. One major bottleneck observed in transformers is learning these long-range dependencies. A couple of works --namely, Longformer \cite{beltagy2020longformer} and Linformer \cite{wang2020linformer}, propose approaches for long-range modeling inputs beyond this limit. Since a function could sometimes be a few hundred lines, these models still fail to capture these long-range dependencies.



In this work, we propose a semantic vulnerability graph (SVG) featuring a rich set of edges capturing semantic and syntactic information, including our novel poacher flow edges to address a variable's information and long-range dependencies \color{black} during execution time. SVG integrates sequential flow for syntactic understanding of the program, data flow to capture how data flows among variables, and a control flow Graph to capture the general flow of statements. \color{black}Poacher flow edges allow the integration of semantic information of source code like long-range dependencies, out-of-scope use of a variable, and divide-by-zero vulnerabilities, by generating extra edges between variables that can potentially bridge the gap between static and dynamic program analysis.\color{black}We propose a transformer and graph neural network-based vulnerability classifier dubbed \textit{Multitask RoBERTa-PFGCN} that includes a large-scale pre-trained RoBERTa \cite{guo2020graphcodebert} on C/C++ source code and utilized Focal Loss (FL) \cite{lin2017focal} to handle data imbalance issues.\color{black} Moreover, \color{black}we propose our new dataset with real-world example functions. Then, by jointly training RoBERTa and GCN modules, our proposed model learns node embeddings using a large-scale Multitask RoBERTa-PFGCN by propagating edge influence through a graph convolution network. The contributions of this paper can be summed up as follows:
\begin{itemize}
    
    \item We propose a unique set of edges dubbed poacher flow edges, such that each of these edges is associated with a set of vulnerabilities. We defined our semantic vulnerability graph (SVG) representation of source code by unifying our introduced poacher flow edges with control, data, and sequential flow edges for vulnerability classification. To the best of our knowledge, we are the first to propose poacher Flow edges, where edges are associated with a particular set of potential vulnerabilities.
    
    \item We propose a joint interpolated multitask unbiased vulnerability classifier comprising a transformer ”RoBERTa” and graph convolution neural network (GCN) with Poacher Flow (PF) edges called RoBERTa-PFGCN, trained using the Focal Loss function to address data imbalance issues. Additionally, RoBERTa-PFGCN provides description and explanation for addressing each vulnerability category. To complement this, we created a large-scale dataset called Vulnerability Finder (VulF) dataset, which contains vulnerability descriptions related to 40 CWE categories. 
    
    \item  We further investigated the effectiveness of our vulnerability classifier by utilizing our proposed dataset VulF and four real-world, large-scale C/C++ vulnerability datasets, including ReVEAL, FFMpeg+Qemu, D2A and MVD. Our experimental results show that our vulnerability classifier outperforms the state-of-the-art results on vulnerability detection with fewer false negatives and fewer false positives.
\end{itemize}

\section{Related Work}
\label{2_related_work}
Earlier works on source code vulnerability detection prominently rely on rule-based systems. Engler et al. \cite{engler2001bugs} propose a technique to automatically extract rules from source code without prior system knowledge. One such rule would be that the declaration of \textit{spin lock} must be followed by \textit{spin unlock} in a C/C++ code to work flawlessly. The simultaneous occurrence of these two statements takes place 99\% in non-vulnerable code. If these statements do not appear in pairs, it is an indication of a security flaw. Essentially, these systems work by creating a rule template for a system. Based on this hypothesis, the authors implemented six checkers, or rules, to identify bugs in code. Founded on this idea, several static analysis based tools like \textit{Flawfinder} \cite{flawfinder}, \textit{RATS} \cite{rats}, \textit{Cppcheck} \cite{cppcheck}, \textit{Coverity} \cite{coverity}, Infer \cite{infer} have been proposed, built on a set of predefined rules to cover a wide range of code vulnerabilities. Since these are rule-based, the rules of these static analyzers need to be updated when a new vulnerability arises and these tools are affected by high false-positive and false-negative rates \cite{yamaguchi2015pattern}.

The work presented by Lin et al. \cite{lin2020software} demonstrates how traditional machine learning (ML) methods offer an alternative to automated vulnerability discovery. In contrast to ML-based vulnerability detection, deep learning-based techniques \cite{dam2018automatic} offer additional possibilities and generalizability. VulDeepecker  \cite{li2018vuldeepecker} proposed detecting vulnerabilities using Bi-LSTM and pre-processed source code by generating \textit{Code Gadget}. According to the authors, a \color{black} \textit{Code Gadget} \color{black} is a collection of data and control dependency statements. $\mu$VulDeepecker \cite{zou2019mu} proposed a multiclass vulnerability classification method using Bi-LSTM. They classified 40 types of vulnerability, with each type tied to a CWE \cite{cwe}. \color{black} Furthermore, Russell et al. \cite{russell2018automated} and Li et al. \cite{li2017software} proposed a TextCNN  based approach to detect vulnerabilities from source code. \color{black} Their proposed approach considers each token as a word embedded to feed a Convolutional Neural Network for training and inference. In recent years, machine learning and deep learning techniques have also been used to detect vulnerabilities in IoT devices \cite{al2022idetect}.

Each of these works considered source code as an analog to natural language, with some limitations in capturing the correct representation of a source code. Since source code is more structured and logical, Bilgin et al. \cite{bilgin2020vulnerability} proposed an AST as a representation technique to detect vulnerability using machine learning. In this approach, the code is converted to an AST. Afterward, to keep the structural information of the code intact, the original AST is converted into a binary AST. The binary AST is flattened using BFS with a CNN for feature generation and classification. Several studies  \cite{lin2019software}, \cite{li2021towards}, \cite{dam2018automatic} including SySeVR \cite{li2021sysevr}, proposed a similar AST based approach with the use of LSTM, Bi-LSTM, or BGRU based methods. VulBERTa \cite{hanif2022vulberta}  RoBERTa \cite{liu2019roberta} and \cite{thapa2022transformer} used a transformer-based model to detect vulnerability from source code.

Although these methods consider using AST to capture the syntactical information of a programming language, these are eventually flattened to feed an encoder that yields the desired vulnerability semantic features. Thus, the original graph syntactics are suppressed. To address this issue, Devign \cite{zhou2019devign} proposed using and preserving the structure of Code Property Graphs (CPGs) \cite{yamaguchi2014modeling} a combination of AST, data and control flow graph, and natural code sequence by using a GGNN \cite{ggnn} combined with a 1d CNN layer to generate the final embeddings for classification. Chakraborty et al. \cite{chakraborty2021deep} have proposed a similar method that makes use of CPGs as an input for training a GGNN \cite{ggnn}. ReGVD \cite{nguyen2021regvd} and LineVD \cite{hin2022linevd} proposed a GCN-based technique for vulnerability detection by creating a graph representation of the source code, and GraphCodeBERT \cite{guo2020graphcodebert} as a tokenizer. \color{black} Moreover, VELVET \cite{ding2021velvet} proposed an ensemble RoBERTa and Gated Graph Neural Network to detect vulnerabilities. 
 \color{black} Each of these techniques offer vulnerability detection at a function or file level, which is not ideal from a programmer's perspective. To address this issue, \cite{nguyen2021information}, \cite{inproceedings} and \cite{mirskyvulchecker} proposed a method that identifies statements contributing to a vulnerability in order to achieve finer granularity in locating vulnerabilities.

Generating a proper graph representation of a program is significant for program analysis \cite{allamanis2022graph}.  Other works found in the literature have proposed an AST-based graph representation for code \cite{allamanis2018learning, wang2021syncobert, jiang2021treebert}. Allamanis et al. \cite{allamanis2018learning} make use of data flow edges with the original AST graph representation. Alon et al. \cite{alon2019code2vec} and \cite{alon2018codeseq} have proposed using ASTs with an attention-based network to generate program representation. TYPILUS \cite{allamanis2020typilus} proposed a graph-based representation similar to \cite{allamanis2018learning}, with the addition of some new edges to predict variable type in a dynamic language. \color{black} CodeBERT \cite{feng2020codebert} learns to represent general-purpose representations for programming languages, while GraphCodeBERT \cite{guo2020graphcodebert} and UniXcoder \cite{guo2022unixcoder} proposes AST graph representation techniques for various programming language-related tasks like code-clone detection, code summarizing, and code translation. \color{black}

\color{black}
To the best of our knowledge, the proposed RoBERTa-PFGCN is the first attempt for code graph representation using programming language structure (data flow, control flow, and sequential flow) combined with Poacher Flow edges to bridge the gap between dynamic and static analysis of a code to improve the performance of code vulnerability understanding.
\color{black}

\section{Multitask Vulnerability Definition}
\label{3_multitask_vulnerability}
\begin{figure}[t]
    \centering
    \includegraphics[width=1\columnwidth]{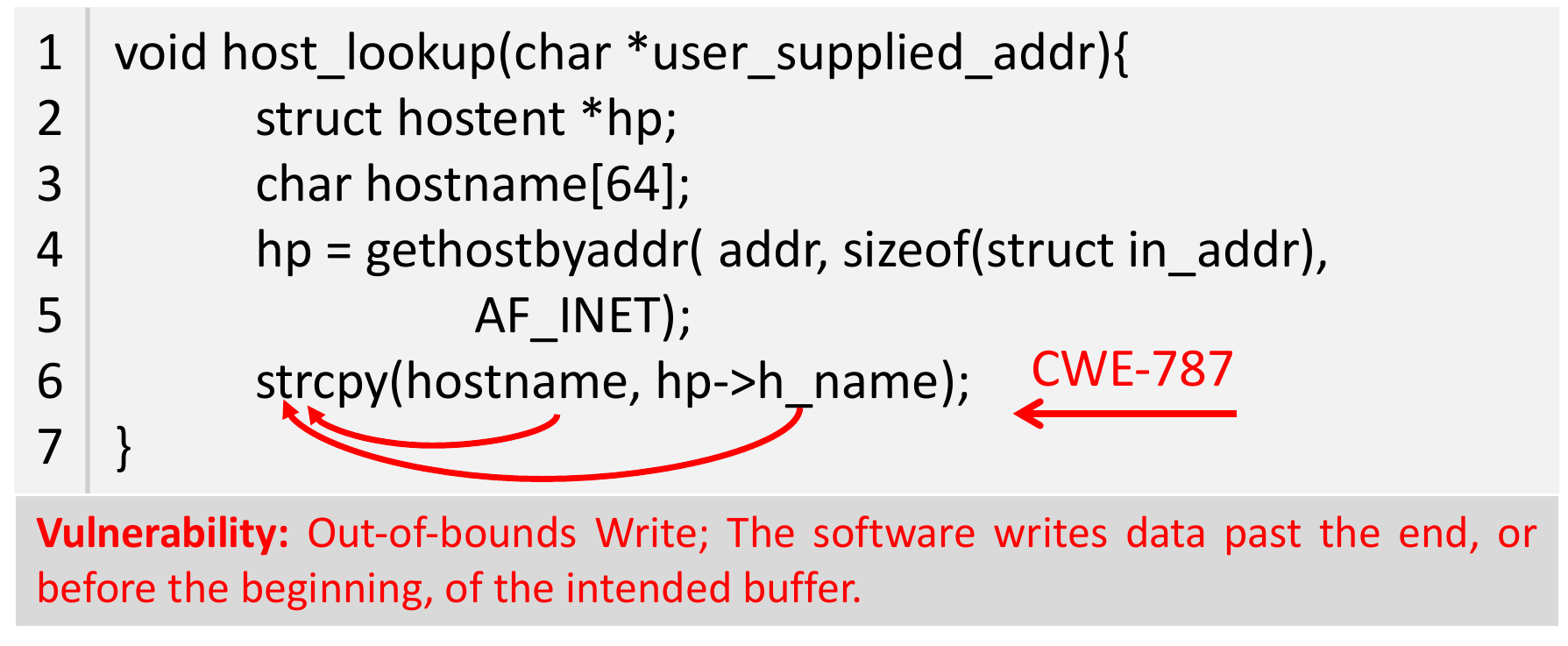}
    \caption{Example of a vulnerability explanation as part of an automated code review process to help developers effectively resolve static software security issues.}
    \label{fig:code}
\end{figure}



\color{black} 
\textbf{Threat Model:} We consider  two types of software developers: (1) adversaries as potential developers who share vulnerable source code or code snippets on online developer collaboration platforms such as StackOverflow, GitHub, or SourceForge; and (2) developers who leverage vulnerable code discovered on online developer collaboration platforms and incorporate it into their software development projects with minor changes. The adversaries can share functions that have possible attack scenarios, such as Remote Code Execution, Buffer Overflow, and Information Leakage to exploit source code vulnerabilities and take control of the system/application, steal data, or launch further attacks. We aim to detect these vulnerabilities early in the development process. Therefore, our approach uses static code analysis to identify code vulnerabilities before execution. Since we are doing a static code analysis, our decision is solely based on the analysis of source code only. Moreover, our system takes external input into consideration during vulnerability analysis. The attack surface may include input function validation, access control, code injection, and configuration management.

\color{black} A variety of vulnerability detectors, such as the work presented in \cite{dam2018automatic}, perform vulnerability classification at a file-level granularity. In contrast, other tool-based approaches, including \textit{Cppcheck} and \textit{Coverity}, depend on a fully compiled or syntactically correct code to run a vulnerability analysis. Despite the advancements proposed in these works, they are language-dependent approaches that cannot be generalized to other programming languages and most of them need a fully compiled program to work correctly. \color{black} We are proposing a generic solution without imposing any constraints on the input language programming function. In  Figure \ref{fig:code} we present a sample function with its output at the bottom. \color{black}

A formal problem definition is presented as follows: a source code function and its label pair are defined as $\{(s_i, v_i)| s_i \in S, v_i \in V \}$ and $i \in \{1, 2, 3, ..., n \}$, where $S$ denotes the set of functions that may or may not be compilable as a standalone program, $s_i$ denotes each function, $n$ denotes total number of functions, and $V = \{ 0, 1 \}$ denotes the set of labels corresponding to each function, where \color{black} the subset of \color{black} vulnerable code is labeled as $V=1$ and \color{black} the subset of \color{black} non-vulnerable code is labeled as $V=0$.

We further extended \color{black} the functionality of our model \color{black} to provide a \color{black}vulnerability classification to the developer. Thus, we define our function and classification pair as $\{(s_i, c_k)\}$ such that $c_k \in C $ and $k \in \{1, 2, 3, ..., n \}$. $C = \{ c_1, c_2, ... c_k \}$ denotes the set of description of the vulnerable function $s_i$, where $k$ is the total number of descriptions that our vulnerability classifier can provide.\color{black} Finally, a pre-processing pipeline converts functions $s_i$ into embeddings using a token embedding generator $\hat{E_R}$ such that,

\begin{figure*}[ht!]
    \centering
    \includegraphics[width=1.0\textwidth]{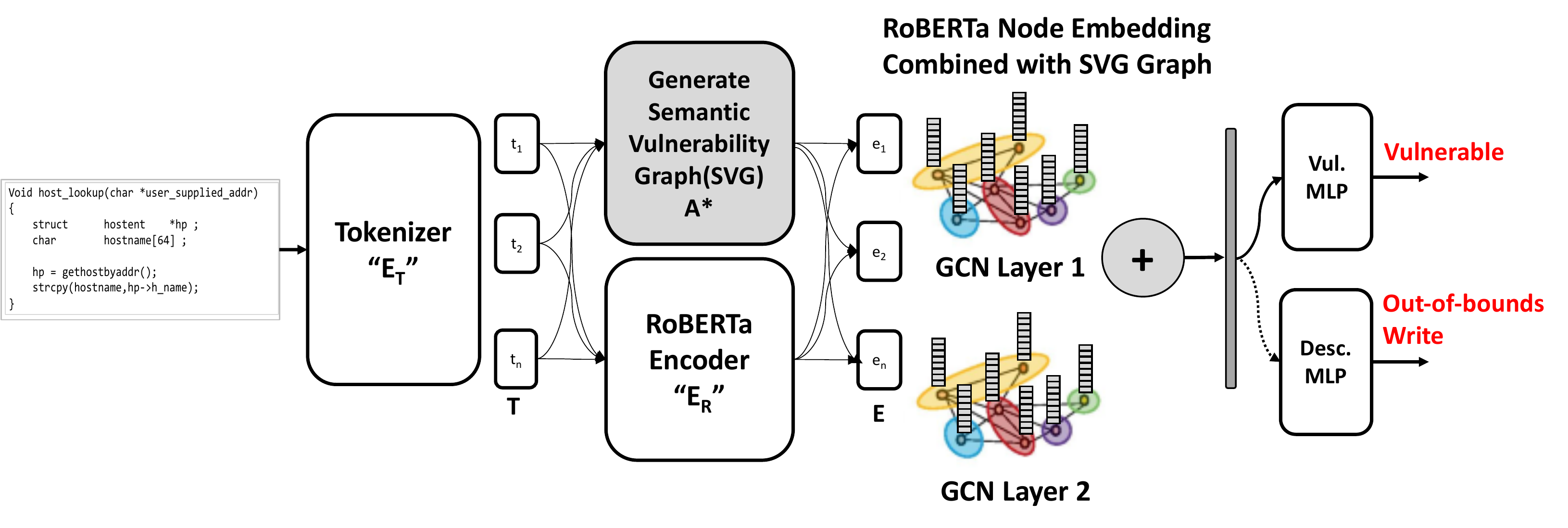}
    \vspace{-0.3cm}
    \caption{Overall Architecture of our Classifier: Our classifier is divided into three parts. Initially, the input source code is pre-processed by creating an SVG. Then RoBERTa layer generates embedding for each token/node of the graph. Finally, the GCN layer takes the node embedding and adjacency matrix for feature generation. Focal Loss forces the model to learn more about the minority class. The MLP layer decides whether a function is vulnerable by leveraging the Focal Loss Function.
  }
    \label{fig:system_algorithm}
    \vspace{-3mm}
\end{figure*}

\begin{equation}
\label{eqn:1}
    e_i = \hat{E_R} (s_i)
\end{equation}

Thus, the set of all embedded tokens of a code is defined as $E = \{e_1, e_2, ..., e_n\}$. We map embedding $E$ with vulnerable code samples $V$, and \color{black}classification $C$ \color{black} such that, $h\textsubscript{1} : E \Rightarrow V $ and $h\textsubscript{2} : E \Rightarrow D $. Therefore, our vulnerability classification $h$\textsubscript{1} function with loss function $\mathcal{L\textsubscript{1}(.)}$ is formally defined as:

\begin{equation}
\label{eqn:2}
\mathcal{L}\textsubscript{1} = \sum_{i=1}^{n}  (h\textsubscript{1} (\hat{E_R} (s_i), v_i | s_i)) 
\end{equation}

Similarly, we define our vulnerability description function $h$\textsubscript{2} with loss $\mathcal{L\textsubscript{2}(.)}$ :


\color{black}
\begin{equation}
\label{eqn:4} 
\mathcal{L}\textsubscript{2} = \sum_{i=1}^{n}  (h\textsubscript{2} (\hat{E_R} (s_i), c_i | s_i)) 
\end{equation}
\color{black}
The overall loss function stands as: 

\[\mathcal{L} = \mathcal{L}\textsubscript{1} + \mathcal{L}\textsubscript{2} + \lambda \frac{1}{2} || w_i ||^2
\]

Our multitask vulnerability detection function learns to detect and provide vulnerability classification by minimizing the loss $\mathcal{L}$. Here, $w_i$ is an adjustable weight learned during training and $\lambda$ is a regularization hyperparameter \cite{ng2004feature}.

In our work, we broadly address the following three Research Questions (RQs):

\textbf{RQ1:} Based on our proposed SVG representation,
can the classifier learn to identify and provide \color{black} CWE Numbers \color{black} of vulnerabilities in real-world source
code?
    
\textbf{RQ2:} Can our classifier learn vulnerabilities in a biased setting?
    
\textbf{RQ3:} Is our classifier generalized enough to detect vulnerabilities in \color{black} N-day and zero-day program samples\color{black}?

\section{Methodology}
\label{4_methodology}
Current graph-based models like CPG and AST generated by tools such as Joern \cite{joern} provide a significant amount of information to detect vulnerabilities in a program. However, runtime vulnerabilities may arise due to the dynamic behavior of program during execution and assignments. Figure \ref{fig:code} depicts a declaration of a variable \textit{hostname} (line 3) and its usage (line 6). Although CPGs provide sufficient information regarding the token dependencies of a graph through data flow, there is no guarantee that the hostname in this case won't be longer than 64 bytes. Furthermore, training a transformer-based model with token sequences from source code is limited given that: 1) code follows a strict syntactic structure compared to the structures found in natural languages, 2) a code's execution time output may produce different behaviors for different input and memory states, and 3) long-range dependencies are commonly found in source code. 

In order to address these problems, our proposed architecture is composed of three main modules, namely: 1) Semantic Vulnerability Graph of a software program, 2) SVG Node Embedding using RoBERTa, and 3) Multitask RoBERTa-PFGCN. Figure \ref{fig:system_algorithm} provides an overall architecture of all the mentioned components.

\subsection{Semantic Vulnerability Graph of a Program}

Our proposed graph representation of a program is denoted as Semantic Vulnerability Graph (SVG). Our SVG is produced via an aggregation of sequential flow edges, control flow edges, data flow edges, and poacher flow edges, a novel edge representing a vulnerability relationship that provides richer information for capturing vulnerability. Each aforementioned element is derived from the same source code. The remaining parts of this subsection provide detailed information on each component used to generate the SVG.

\paragraph{\textbf{Node Generation Using Tokenizer}} A token is a series of characters separated by spaces or punctuation marks generated by a tokenizer. Tokens may take the form of words, integers, real numbers, or a combination of these. However, tokens differ slightly when they are used in Programming Language Processing problem. In programming language, tokens may come in the form of \textit{camelCasing} or \textit{snake\_casing}. Consider an example of the token \textit{get\_item}. In Natural Language Processing, the tokenizer will separate the word into two tokens, \textit{get} and \textit{item}. However, this combination is treated as a single token since the input is a code. Moreover, other symbols (such as parenthesis, semicolons, etc.) are considered as a single token. Each of these tokens are used as a node of our SVG. \color{black} When we tokenize our code, it generates three features for each token, the original token itselt, its position, and the token type.\color{black}

\paragraph{\textbf{Adjacency Matrix Definition}} Let us consider that a graph has an adjacency matrix $A$ where $m$ and $n$ are some arbitrary nodes in the graph and edges are the connection between two nodes. Thus, our adjacency matrix is defined as:

\[
A_{m, n} = \left\{\begin{matrix}

1 & if & edge & exists \\
0 & & Otherwise
\end{matrix}\right.
\]

where $A=1$ indicates an edge exists between two arbitrary nodes $m$ and $n$ and $A=0$ indicates otherwise.

\paragraph{\textbf{Data Flow Edges}} Show the usage and modification of a variable \cite{zhou2019devign}. Data flow edges are defined as a connection between two variables dependent on each other during value assignment or modification or other usage. Some other usage of the variable may include variable definition, initialization, update, or alteration.

\paragraph{\textbf{Control Flow Edges}} Illustrate the statements or operations executed throughout the program \cite{zhou2019devign}. The alternate execution of statements may be determined by conditional statements (e.g., if/while/switch).

\paragraph{\textbf{Sequential Flow Edges}} Demonstrate the syntactic relationship between the tokens of a program inspired from \cite{huang2019text, zhang2020every}. Sequential flow edges show the connection of a token with its neighboring tokens. To generate this edge, we create an edge from a token with its subsequent neighboring tokens. The number of subsequent tokens the initial token is connected to is determined during the experiment.


\color{black}
\paragraph{\textbf{Poacher Flow (PF) Edges}}  
We defined Poacher Flow edges to bridge the gap between dynamic and static analysis of source code. As opposed to programming language structure (data flow, control flow, and sequential flow), PF edges are meant to identify program boundaries, potential corner cases, and external checkpoints. This is accomplished by considering the external environment context in which the program operates, including insecure input handling, the use of unsafe functions, SQL injection, or unauthorized code execution that have just recently been discovered by the CWE community in programs of a similar nature. Our goal is to bridge the gap between dynamic and static analysis of a program by using PF edges. Specifically, PF edges serve as a connection between the knowledge and patterns learned stochastically from known existing vulnerability patterns using labeled data by incorporating PF edges into the machine learning training procedure. We have identified three categories of PF edges: data processing edges, access control edges, and resource management edges. Each edge of these edge categories is discussed in detail in the subsections below. In addition, \color{black} Algorithm \ref{alg:algorithm_1} presents the approach for generating all the elements of Poacher Flow Edges. \color{black}

\color{black}

\renewcommand{\algorithmicrequire}{\textbf{Input:}}
\renewcommand{\algorithmicensure}{\textbf{Output:}}
\algnewcommand{\LeftComment}[1]{\Statex \(\triangleright\) #1}
\begin{algorithm}[t]
\caption{\color{black}Generating Poacher Flow Edges\color{black}}
\label{alg:algorithm_1}
\begin{algorithmic}[1]

\Require Source Code
\Ensure Poacher Flow Edges
\Procedure{\textit{POACHER\_EDGES}}{$code$}

\State $tokens \gets Tokenizer(code)$
\State $asst\_operators \gets [=, +=, -=, <<=, ...]$
\State $adj\_matrix \gets [][]$
\State $stack \gets []$

\LeftComment{Dictionary Initialized}
\State $scope \gets \{  \}$  
\For{$token$ in $tokens$}

    \LeftComment { \textbf{Data Processing}}
    \If{$token.type$ in $asst\_operators$}
        \State $left \gets getPrevToken(token)$
        \State $right \gets getNextTokens(token)$
        \State $adj\_matrix[left][right] \gets 1$
    \EndIf
    \If{$token.type$ is $"API"$}
        \State $params \gets getParameters(token)$
        \State $adj\_matrix[token][params] \gets 1$
    \EndIf
    
    \LeftComment { \textbf{Access Control}}
    \State $FunParams \gets getFuncParams()$
    \If{$token.type$ is $"execution"$}
        \If{$token$ is not checked before asst}
            \State $next \gets getNextToken(token)$
            \If{$next$ is subset $FunParams$}
                \State $adj\_matrix[token][next] \gets 1$
            \EndIf
        \EndIf
    \EndIf

    \LeftComment { \textbf{Resource Management}}
    \If{$scope[token]$ is $end$}
        \State  $adj\_matrix["free"][token] \gets 1$
    \EndIf
    \State $stack.push(token)$
    \If{$pairMatch(token)$}
        \State $pair\_token \gets stack.pop()$
    \EndIf
        \If{$token$ is $"free"$}
        \State $next\_token \gets getNextToken(token)$
        \State $scope[next\_token] \gets end$
    \EndIf

\EndFor

\State \Return $adj\_matrix$
\EndProcedure
\end{algorithmic}
\end{algorithm}

\color{black}
\textit{Data Processing Edge:} Data processing vulnerabilities are the most common types during the software development stage. For example, Out-of-Bounds Read is ranked 1 out of the top 25 vulnerabilities from 2022 \cite{cwe}. Data flow edges are useful for capturing the flow of data, but may not be sufficient for capturing complex data  operations, such as memory pointer arithmetic. Additionally, when data manipulation involves APIs such as (\textit{strcpy, read, and write}), data flow edges may fail to capture this information. The data processing edge is an extension to the existing data flow graph, which estimated the potential outcome of various mathematical operations, illegal memory issues, and unsafe API execution, pointer arithmetic. For instance, estimating divide by zero, using an uninitialized variable or using unsafe APIs like \textit{gets()} in C/C++.
\color{black}



\textit{Access Control Edge:}
According to the Open Web Application Security Project (OWASP), software and data integrity failures are ranked among the top ten web application security risks \cite{owasp10}. These attacks take advantage of improper neutralization of special elements in web page output. While programming language structures (data flow, control flow, and sequential flow) cannot discover these vulnerabilities, access control edges can be utilized to address this issue. These edges correspond to external program calls, including application configuration settings that may not be present in the application's source code, such as passing untrusted data as arguments. Other edges include improper control over code generation and improper neutralization of special elements used in SQL commands. By performing conditional edge checks, it is possible to prevent malicious actors from passing untrusted data as arguments.



\textit{Resource Management Edge:} Software vulnerability may occur when resources are not adequately managed including when a buffer copy is executed without verifying input size, incorrect array index validation, resource exhaustion, utilization of memory after an uncontrolled allocation or incomplete cleanup, or incorrect synchronization of resources within an exclusive operation such as semaphore. These scenarios can be captured by Resource management edges, which can make the classifier aware of potential inadequate resource management operation.



\begin{figure*}[ht!]
\begin{minipage}[b]{1\linewidth}
  \centering
  \centerline{\includegraphics[width=0.8\textwidth]{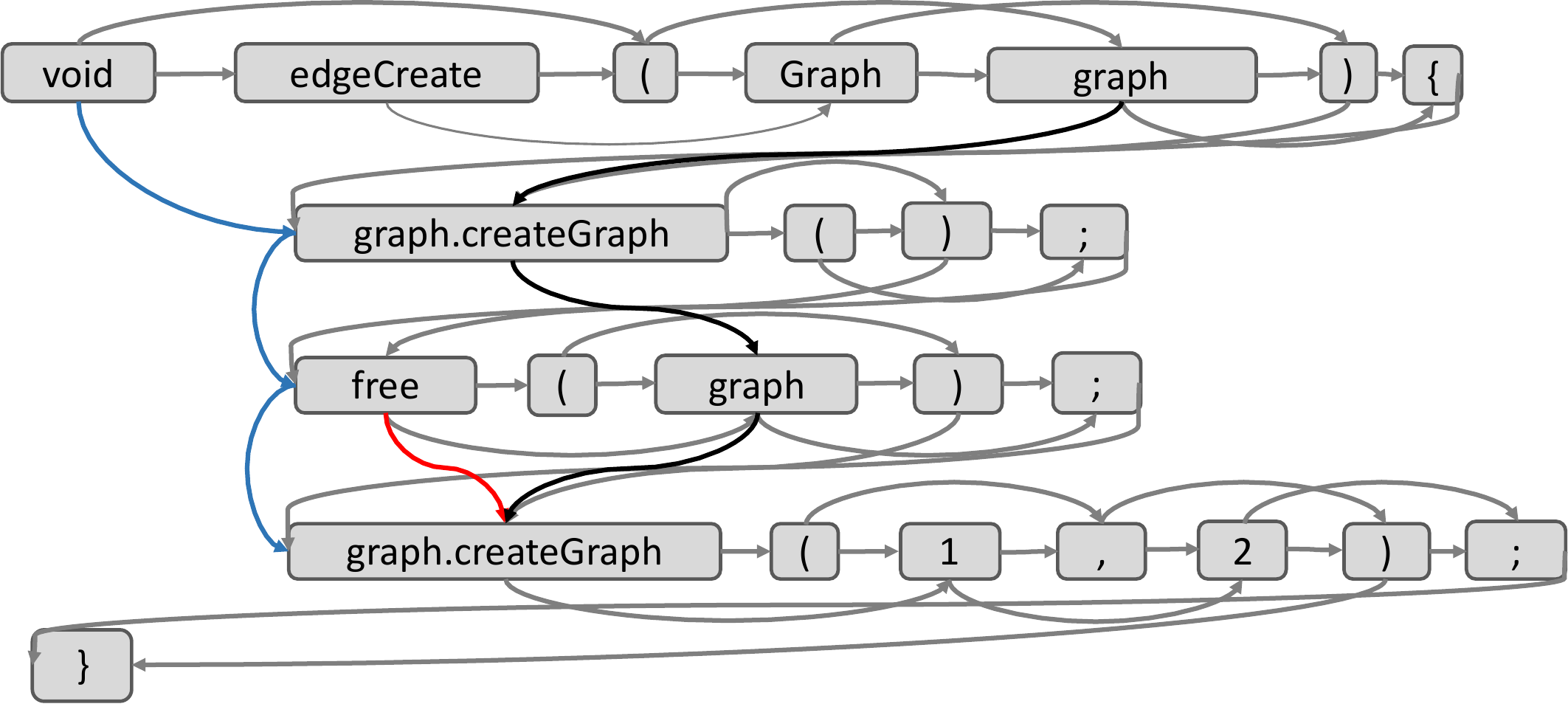}}
  \caption{Depiction of our SVG. Each gray box shows individual tokens of our SVG. The red line depicts a poacher flow edge, the black line depicts data flow edges, the blue line depicts control flow edges and the gray line depicts sequential flow edges. 
  }
  \label{fig:explain}
\end{minipage}
\end{figure*}


\paragraph{\textbf{Combining the Edges as SVG}} Each edge type is critical for finding vulnerabilities in a function. Data flow edges (black edges), shown in Figure \ref{fig:explain} identify the flow of data for each variable; control flow edges (blue edges) are responsible for the overall flow of programs; sequential flow edges (gray edges) shows the syntactic relationship between the tokens of the program; lastly, Poacher Flow edges (red edges) are meant to bridge the gap between dynamic and static analysis of code by generating edges related to program boundaries, corner cases, and external checks. The SVG is constructed through the combination of data flow, control flow, sequential flow, and poacher flow edges. SVG produces richer semantic and syntactic information necessary for vulnerability detection and classification. Figure \ref{fig:explain} presents an example of an SVG composed of 68 edges in total, including 61 sequential flow edges (gray edges), 3 data flow edges (black edges), 3 control flow edges (blue edges), and 1 Poacher Flow edge (red edge).






\color{black}

\subsection{SVG Node Embedding using RoBERTa}

\paragraph{\textbf{SVG Node Embeddings}} RoBERTa \cite{liu2019roberta} is used to generate embeddings for each token in our graph. RoBERTa was built upon BERT \cite{devlin2018bert}, in which the system learns to predict purposefully masked text within unannotated language examples. RoBERTa modifies critical hyperparameters in BERT, such as deleting BERT's next-sentence pre-training target and training with significantly larger minibatch sizes and learning rates. This pre-training technique allows RoBERTa to outperform BERT in terms of the masked language modeling objective and improves the performance of subsequent tasks. However, to tokenize and initialize the node embeddings, a pre-trained variant of RoBERTa presented in GraphCodeBERT \cite{guo2020graphcodebert} is used for source code representation on C/C++. The classifier makes use of word embeddings generated by RoBERTa and embeddings generated by a GCN model fed with heterogeneous SVG. The embeddings $E$ generated by the pre-trained RoBERTa encoder are as follows:

\[
E = \hat{E_R} (T)
\]
Here, the set of tokens $T$, where $t_i \in T, i = 1, 2,  ..., n$, is the set of $n$ tokens in SVG that are used as the input for the RoBERTa encoder $\hat{E_R}$.

Afterward, an adjacency matrix $A_{m, n}$ is created by using the set of tokens, $T$, and the connections observed between tokens following our SVG. After this step, the adjacency matrix is converted into a heterogeneous multi-edged graph $\mathbb{G} (T, E, A)$, where $E \in \mathbb{R} \textsuperscript{\textit{d}} $ is the $d$ dimensional embedding or feature vector of each token $t$ in the graph. While different edge types compose our SVG, only a single adjacency matrix is used to represent all the edges where a value of 1 is set if any of the edges exist between two tokens and 0 if the edges do not exist.

\subsection{Multitask RoBERTa-PFGCN}

In an SVG, the existence of specific edges could serve as indicators of the existence of vulnerability. Graph convolution networks (GCN) are designed to comprehend the edge connection between two nodes. GCN is used to capture the relationship between the elements of $\mathbb{G} (T, E, A)$ that are essential for vulnerability detection.
GCN is composed of two layers that aggregate vector representations of a node from its neighbors with a residual connection. GCN is formulated as follows:

\begin{equation}
\label{eqn:5}
H\textsuperscript{(n + 1)} = \sigma \biggl( W^n H\textsuperscript{n} A^* \biggr )
\end{equation}

where $W^n$ represents the weights at $n$-th layer during training and H\textsuperscript{n} is the feature representation of nodes at $n$-th layer. Thus, $H\textsuperscript{(0)} = E$ while $A^*$ is the normalized adjacency matrix. Matrix multiplication is done on $W^n$, $H^n$, and $A^*$, which goes through an activation function $\sigma$ (e.g., $ReLU$).

The values in the adjacency matrix $A$ are normalized to prevent numerical instabilities, such as vanishing or exploding gradients, that might prevent the model from converging into an optimal solution.  The adjacency matrix is normalized using the method proposed by Kipf et al. \cite{kipf2016semi}, which performs an inverse dot product operation for normalization. Let us consider $\hat{D}$ as the diagonal node degree matrix such that, $\hat{D}\textsubscript{ij} = \sum_{j}^{} A\textsubscript{ij}$. The degree matrix of a graph is a diagonal matrix that records the degree of each vertex or the number of edges that connect each vertex to another vertex. $\hat{D}$ also contains information about the number of edges attached to each vertex. The normalized adjacency matrix is computed as:


\[
A^* = \hat{D}\textsuperscript{-1} . A  \\
\]
which is equivalent to:

\[
A^* = \hat{D}^\frac{-1}{2} . A . \hat{D}^\frac{1}{2}
\]
According to the authors in \cite{kipf2016semi}, the latter formula is used for better normalization. 


\paragraph{\textbf{Residual Connection}} In the work presented by He et al. \cite{he2016deep}, a residual connection is used to propagate feature representation learned from the $(H^n)$  layer to the next layer $(H\textsuperscript{n+1})$ by allowing gradients  to pass directly from one layer to the next without encountering a vanishing or exploding gradient problem. By adding the residual connection, our model is redefined from Equation \ref{eqn:5}  as:

\begin{equation}
\label{eqn:6}
H\textsuperscript{(n + 1)} = H\textsuperscript{n} + \sigma \biggl( W^n H\textsuperscript{n} A^* \biggr )
\end{equation}

After this, two dense parallel layers are added. The first layer consists of two neurons that provide the outcome for vulnerability detection. The second layer consists of 41 neurons that indicated the vulnerability description associated with the detected vulnerability.

\paragraph{\textbf{Loss Function}} Vulnerability in a real-world setting appears highly imbalanced. As a consequence, non-vulnerable code highly outnumbers vulnerable code, thus a classifier is always biased toward the majority class. As a result, usual loss function like CrossEntropyLoss provides higher false-positive and false-negatives.  We employ \textit{Focal Loss} \cite{lin2017focal} to rectify the class imbalances of our datasets. Without this approach, our model would learn biases towards non-vulnerable samples, drastically affecting our classification performance. The Focal Loss is denoted based on cross-entropy (CE) loss for binary classification problems as:

\[
CE{p, y} = \left\{\begin{matrix}
-\log(p) & if & y=1 \\
-\log(1-p) & otherwise,
\end{matrix}\right.
\]
where $y=\{0, 1\}$ denotes the ground truth provided to the classifier during the training process and  $p=\{0, 1\}$ is the models output probability for the class $y=1$, for binary classification. However, we expanded this for multitask classification as well. For convenience, probability distribution $p_t$ is defined as:

\[
p_t = \left\{\begin{matrix}
p & if & y=1 \\
(1-p) & otherwise,
\end{matrix}\right.
\]
Focal Loss integrates a weighing factor $\alpha \in [0, 1]$ and defines the mathematical expression of Focal Loss for a binary classification problem. Thus, the balanced CE loss can be rewritten as:

\begin{equation}
\label{eqn:7}    
CE(p_t) = -\alpha \log(p_t)
\end{equation}

Vulnerability classification without the loss function shows that the classifier can be confused by the majority class, which also dominates the gradients. Although $\alpha$ balances majority and minority examples, it does not differentiate between easy (positives/negatives samples that are predicted as positive/negative) and hard examples (positives/negatives samples that are misclassified as negative/positive). To overcome this issue, a modulating factor $\delta$ is used with the cross-entropy loss to down-weight easy examples, which forces the model to be trained more precisely on hard negatives. By combining weight balance and Focal Loss, our final Focal Loss function from Equation \ref{eqn:7} is defined as follows:

\begin{equation}
\label{eqn:8}    
FocalLoss(p_t) = -\alpha (1 - p_t)^\delta \log(p_t)
\end{equation}

Where, $\gamma$ is an adjustable parameter and $\gamma \geq 0$. Figure \ref{fig:explain} shows the overall architecture of our proposed vulnerability classifier.

\paragraph{\textbf{Complexity analysis}} Algorithm \ref{alg:algorithm_1} provides the order of logics to create our graph. Given a sample code as input, RoBERTa Transformer is used to tokenize the code snippet to generate $n$ tokens. Thus, the time complexity to generate $n$ tokens is $O(n)$. In order to generate each Poacher Flow edge, the tokens are iterated once and all edges are created in a single pass. We generated data flow, control flow, and sequential flow edges in a single pass by iterating over $n$ tokens, hence the time complexity is $O(n)$.  As a result, the overall time complexity to generate our complete SVG is $O(n)$. However, other graph-based analysis \cite{guo2022unixcoder} \cite{guo2020graphcodebert} \cite{zhou2019devign} consist of generating an AST, which can be very time consuming. For example, for the same program with $n$ tokens, the time complexity to insert a single token into an AST is $O(\log n)$ on an average case when the tree is balanced. However, when the tree is imbalanced, the time complexity to insert a single element is $O(n)$. Thus, the time complexity to generate an AST by inserting $n$ elements in a balanced tree is $O(n \log n)$ and in an imbalanced tree is $O(n^2)$, which is much higher than our proposed SVG.

\section{Experiments And Discussions}
\label{5_experiments_discussion}
\begin{table*}[ht]
\caption{Summary of our VulF dataset with total number of functions for each CWE labels with \color{black} short \color{black} description.}
\label{tab:multiclassvuln}
\begin{tabular}{@{}lclc@{}}
\toprule
\multicolumn{1}{c}{CWE Short Description}          & \# of Functions & \multicolumn{1}{c}{CWE Short Description}                & \# of Functions \\ \hline
Non-Vulnerable-N/A                                 & 115550          & 467 - Using of null pointer                              & 508             \\
020 - Process data without validation              & 70              & 469 - Incorrectly determining pointers size              & 1701            \\
020, 665, 400 - Consuming resource without control & 365             & 476 - Trying to access a dereferenced pointer            & 421             \\
074 - Injection of foreign code                    & 1640            & 506 - Containing malicious code                          & 102             \\
119 - Performing read/write outside buffer         & 4713            & 573 - Calling an API incorrectly                         & 221             \\
119, 666, 573 - Operation outside memory buffer     & 305             & 662, 573 - Calling API without sync.                     & 331             \\
138 - Unchecked use of special elements            & 200             & 573, 666 - Incorrectly following configuration           & 306             \\
170 - Incorrectly terminating a string             & 1006            & 610 - Use of externally controlled resource              & 309             \\
187 - Incorrect comparison of string               & 506             & 662 - Incorrect synchronization of thread                & 307             \\
190 - Integer overflow by mathematical operation   & 326             & 665 - Incorrect initialization of a resource             & 305             \\
191 - Integer underflow by mathemetical operation  & 68              & 666 - Performing operation on resource in wrong lifetime & 734             \\
221 - Misinterpret records                         & 60              & 668 - Exposing resource to incorrect sphere              & 805             \\
311 - Product does not encrypt critical data       & 581             & 670 - Malicious Incorrect Control Flow                   & 601             \\
327 - Use of risky algorithm or protocol           & 35              & 673 - Changing control sphere by external party          & 61              \\
362 - Concurrent execution of shared resources     & 211             & 676 - Use of dangerous function                          & 1380            \\
369 - Division ob zero error                       & 289             & 704 - Incorrectly converting type of a resource          & 480             \\
400 - Consuming resource without limit             & 107             & 706 - Accessing incorrectly resolved resource            & 206             \\
400, 404 - Release consuming resource              & 1200            & 754 - Improper check on unusual exceptions               & 78              \\
400, 665 - Consuming uninitialized resource        & 1560            & 758 - Improperly use of API                              & 59              \\
404 - Release a resource incorrectly               & 508             & 834 - Iteration of a loop eccessively                    & 210             \\
668, 404 - Improper use of Resource                & 2860            &                                                          &                 \\ \hline
                                                   &                 & \multicolumn{1}{c}{}                                     & Total = 141285  \\ \hline

\end{tabular}
\end{table*}

\color{black}
Our experiments were conducted using highly balanced, mildly unbalanced, and highly unbalanced datasets, as well as real-world N-day and zero-day program samples collected from publicly available resources for evaluation purposes. Our experiment was designed to evaluate three metrics in mind: our model's ability to classify vulnerabilities with their corresponding descriptions, our model's ability to handle a biased dataset, and the importance of detecting N-day and zero-day programs.\color{black}


We tested our model's vulnerability classification across various experimental settings in which each experimental subset was chosen to resolve its respective Research Question (presented in Section \ref{3_multitask_vulnerability}). In addition, we provided an ablation study to observe our model's performance \color{black} by adding sub-components of PF edges to observe their impact on vulnerability detection. The remaining content of this section is divided into the following subsections: Datasets, Data Pre-Processing, Performance Evaluation, Time and Memory analysis, and Ablation Studies.\color{black}

\subsection{Datasets}

We utilized different datasets that included highly balanced, mildly unbalanced, and highly unbalanced data. Particularly, we utilized the large-scale MVD \cite{zou2019mu} dataset since it includes both vulnerability data and a CWE number for each source code function. This dataset is comprised of a huge number of real-world and synthetic vulnerability samples, and it is mildly unbalanced. FFMpeg+Qemu \cite{zhou2019devign} and D2A \cite{zheng2021d2a} are two more balanced real-world datasets we have included. We also utilized the ReVEAL \cite{chakraborty2021deep} dataset, a highly unbalanced real-world dataset with a non-vulnerable to vulnerable data sample ratio of 9:1. Lastly, we created VulF by aggregating publicly accessible source code from GitHub and the National Vulnerability Database. In addition, we used samples from other existing datasets to build a wild dataset that accommodates the necessary subpopulation shift \cite{koh2021wilds}, \cite{santurkar2020breeds} for detecting real-world vulnerabilities. Table \ref{tab:datacomparison} provides a brief description of the datasets.

\paragraph{\textbf{Vulnerability Finder (VulF)}} We created VulF, a large-scale dataset comprised of data from multiple publicly available sources. We started Vulf from data collected from the National Vulnerability Database \cite{nvd}, consisting of vulnerable and non-vulnerable code samples. Each vulnerable source code was mapped to a Common Weakness Enumeration (CWE) number. A list of the most prevalent vulnerability categories was built; yet, only source code samples written in C/C++ were kept in our dataset. For example, CWE-119, Improper Restriction of Operations within the Bounds of a Memory Buffer, defines a software vulnerability that occurs when software reads or writes data past the specified buffer's limit or after its specified size. CWE-020 (Improper Input Validation), another top vulnerability, refers to a program that accepts input or data but fails to validate it before use. As a result, an altered control flow, arbitrary control of a resource, or execution of arbitrary code may occur. Additionally, we combined data from $\mu$VulDeePecker's MVD dataset\cite{zou2019mu}. In order to make our dataset more robust, we implemented Code Reformatting, Beautification, Dead Code Elimination, Variable Renaming, Identifier Mangling, and Dead Code Insertion techniques presented by Jain et al. \cite{jain2020contrastive} to generate synthetic data.

Furthermore, we enhanced our Vulf dataset by generating descriptions linked to each CWE. To ensure the usefulness of our descriptions from developers' perspective, we engaged a group of junior programmers with limited knowledge of software security vulnerabilities. They validated the effectiveness and usefulness of our CWE descriptions for vulnerability root cause analysis and code fixes. Their feedback helped us refine our CWE descriptions to ensure they are useful and effective for developers seeking to fix source code vulnerabilities. The collected data was divided into 40 vulnerability categories and one benign category. Each vulnerable function was labeled with its corresponding CWE number obtained from\cite{cwe} and description. We present further information regarding our dataset in Table \ref{tab:multiclassvuln}.

\paragraph{\textbf{ReVEAL \cite{chakraborty2021deep}}}
The ReVEAL dataset was curated by Chakraborty et al. \cite{chakraborty2021deep} by tracking vulnerabilities in two open-source projects: Chromium and Linux Debian. Chromium is an open-source project of Chrome. The authors crawled Bugzilla and Linux Debian Kernel via the Debian Security Tracker to generate their dataset. This dataset reflects the 9:1 ratio of vulnerable to benign code described in Table \ref{tab:datacomparison}.

    
\paragraph{\textbf{FFMpeg+Qemu \cite{zhou2019devign}}}
The FFMpeg+Qemu dataset is a collection of real-world source code vulnerability detection data compiled by Devign. \cite{zhou2019devign}. This dataset contains four repositories, Linux Kernel, Qemu, Wireshark and FFMpeg, but the authors only share the FFMpeg and Qemu datasets publicly. Their data annotation was conducted using the Commit Filtering approach proposed by \cite{zhou2017automated}, and manual \color{black} verification \color{black} was completed by four experienced security researchers for final verification, devoting 600 person-hours to the task.
    
\paragraph{\textbf{D2A \cite{zheng2021d2a}}}
IBM Research has assembled a real-world vulnerability dataset, D2A \cite{zheng2021d2a}. They included open-source projects like FFMpeg, OpenSSL, httpd, NGINX, libtiff, and libav in this dataset. This dataset was created using a differential analysis-based method, wherein the authors initially \color{black}filtered the commit messages \color{black} to sort out potentially vulnerable commits before using three static analyzer tools, \textit{CppCheck}, \textit{FlawFinder}, \textit{Clang Static Analyzer}, and \textit{Infer}, for a two-way checking.

\color{black}
\paragraph{\textbf{MVD \cite{zou2019mu}}}
MVD curated by Zou et al. \cite{zou2019mu} is a multiclass vulnerability dataset of 40 vulnerable classes and one benign class. The datasets were collected from NIST \cite{nist}, and SARD \cite{sard}. A significant number of program samples of this dataset consist of synthetic, vulnerable, and non-vulnerable code examples. Table \ref{tab:datacomparison} briefly describes the number of benign and vulnerable, their ratio, and the total number of code samples for each dataset.


\color{black}

\begin{table}[b]
\centering
\caption{\color{black}Overview of datasets utilized for training and testing, encompassing highly balanced, mildly unbalanced, and highly unbalanced sets.\color{black}}
\label{tab:datacomparison}
\begin{tabular}{lllll}
\toprule
\textbf{Data Source} & \textbf{Benign} & \textbf{Vulnerable}  & \textbf{Total}  & \textbf{Ratio} \\
\midrule
D2A \cite{zheng2021d2a}           & 3222           & 3506               & 6728  & $\sim$1:1.08   \\
FFMpeg + Qemu \cite{zhou2019devign} & 14854          & 12460              & 27314 & $\sim$1.19:1   \\

MVD \cite{zou2019mu}        & 138522          & 43119               & 181641 & $\sim$3.2:1      \\

VulF (Ours) & 115550     & 25735 &  141285 &     $\sim$4.5:1 \\

ReVEAL \cite{chakraborty2021deep}        & 20494          & 2240               & 22734 & $\sim$9:1      \\
\bottomrule
\end{tabular}
\end{table}

\subsection{Data Pre-Processing}

\paragraph{\textbf{Graph Data Preparation}}
After \color{black} collecting \color{black}the datasets, they were converted into an SVG for further analysis. To convert the program into a graph, we add a starting \textit{\textlangle{}s\textrangle{}} and ending token \textit{\textlangle{}\textbackslash s\textrangle{}} at the beginning and end of each program. Then we convert the code into a sequence of tokens using a RoBERTa tokenizer, pre-trained on C/C++ programs for code representation. Next, we convert the sequence of tokens into SVG which  a sequential flow, data flow, control flow, and poacher flow edges are generated. Finally, Algorithm \ref{alg:algorithm_1} creates an adjacency matrix of shape $n \times n$, where $n$ is the total number of tokens in the graph. As a second step in creating the graph, each node of our SVG was encoded by generating a word embedding of size 768 using pretrained RoBERTa.

\begin{table}[b]
\centering
\caption{Three sample codes were randomly selected from our VulF dataset to show how the number of nodes and edges of AST compares with our graph.}
\label{tab:datasource}
\begin{tabular}{l|cc|cc|c}
\toprule
Code Sample    & \multicolumn{2}{l|}{SVG (Our Graph)}   & \multicolumn{2}{l|}{AST Graph}   & Edge Ratio \\
\midrule
               & \multicolumn{1}{l|}{Node} & Edge & \multicolumn{1}{l|}{Node} & Edge &  \\
\midrule

Example 1 & \multicolumn{1}{l|}{3660} & 8516 & \multicolumn{1}{l|}{2875} & 5748 &  1.5:1\\             
Example 2 & \multicolumn{1}{l|}{3324} & 6640 & \multicolumn{1}{l|}{2415} & 4828 &  1.4:1\\              
Example 3 & \multicolumn{1}{l|}{4056} & 8114 & \multicolumn{1}{l|}{3909} & 7816 &  1:2.5\\            

\bottomrule
\end{tabular}
\end{table}

\paragraph{\textbf{SVG Analysis}} Table \ref{tab:datasource} presents a comparison between the number of edges and nodes generated by SVG to those generated by an AST in three different sample codes. These code samples were selected randomly from the VulF dataset. As can be observed, SVG generates more edges than an AST. During the generation of the AST, we observed that several intermediate nodes were generated which removed parentheses. Due to this, the edges coming in and out of these intermediate nodes of AST contain no information about the code snippet's vulnerability, regarding these edges useless. In comparison, our SVG retains the parentheses, semicolons, and all other symbols of a programming language as nodes within the graph. Given this additional relationship between tokens, our classifier can attain a higher comprehension of the code snippet semantically and syntactically.

\subsection{Performance Evaluation}
We randomly split all datasets with a ratio of 80:10:10 for training, validation, and testing. We implemented RoBERTA-PFGCN with a 12-layer RoBERTa encoder to generate the token embeddings. 
\color{black} We also completed a time and memory analysis using RoBERTa-AST, in which we generated an AST instead of SVG for our GCN. To generate features, the tokens were converted into their equivalent SVG representation. We created a two-layer graph convolutional neural network with a residual connection from the first layer's input to the second layer's input. The dimensionality of the hidden layer is set to 128, with a learning rate of 5e-4 when training a balanced dataset. When we used ReVEAL, the highly unbalanced dataset, we used 1e-5 as the learning rate and trained the model for 100 epochs with a batch size of 512, while the maximum token length was set to 400. We trained our model on 8 DGX-A100 NVIDIA GPU, wherein each model training \color{black}and testing \color{black} session took 4-6 hours to complete due to the data size.

\paragraph{\textbf{Evaluation Metrics.}}
Our work was evaluated using four metrics: Accuracy, Precision, Recall, and F1. Our model's predictions were categorized as True Negative (TN), True Positive (TP), False Negative (FN), and False Positive (FP). TP refers to samples correctly classified as vulnerable, FP to samples incorrectly classified as vulnerable, TN to samples correctly classified as benign, and FN to samples incorrectly classified as benign. Using these statistics, we compute the Precision as $P = \frac{TP}{TP + FP}$, Recall as, $Recall = \frac{TP}{TP + FN}$, and the F1 score as,\vspace{1mm} $F1 = 2 \times \frac{Precision * Recall}{Precision + Recall}$.\vspace{1mm}

The remainder of this section will outline experiments designed and performed to explore each research question.
\\\\
\textbf{RQ1: Based on our proposed SVG representation,
can the classifier learn to identify and provide \color{black} CWE Numbers \color{black} of vulnerabilities in real-world source
code?}

\begin{figure*}[t]
        \centering
        
        \includegraphics[width=0.9\textwidth]{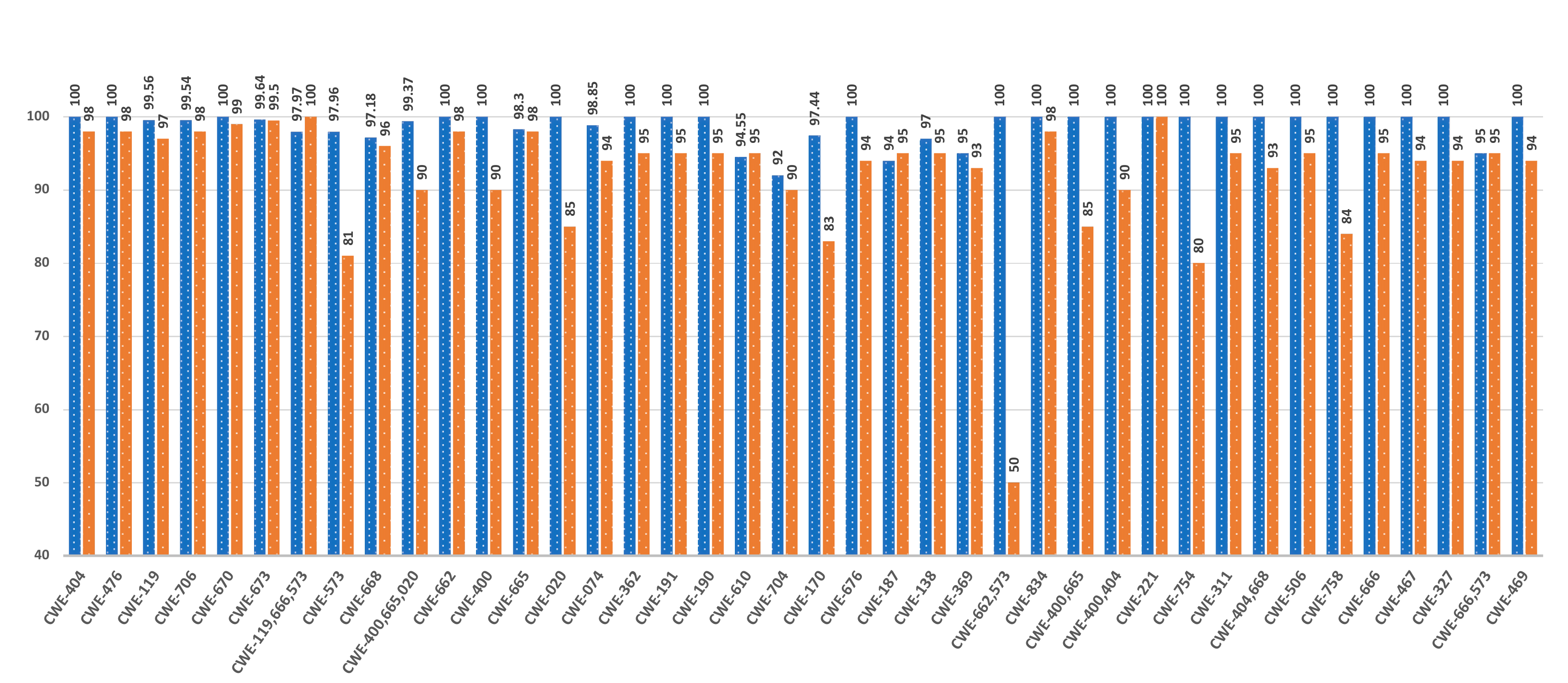}
        
    \caption{\color{black}
    CWE class-by-class F1 score comparison on our proposed Model (RoBERTa-PFGCN), vs. $\mu$VulDeepecker on MVD dataset provided by $\mu$VulDeePecker including 40 CWE classes. The blue bar corresponds to RoBERTa-PFGCN, while the orange bar represents $\mu$VulDeepecker.\color{black}}
    \label{fig:40_class}
\end{figure*}

A vulnerability classification system should generate semantic features to detect vulnerable patterns from source code and classify vulnerabilities with higher accuracy and lower false positive and false negative rates. To test the effectiveness of our classifier, it was trained and evaluated with the five datasets mentioned earlier.

In order to test how our vulnerability \color{black}model detect and classify \color{black}each type of vulnerability, our model was tasked with providing a CWE number \color{black} and a CWE description \color{black} for each type of vulnerability. As a result, the software developer can properly understand the detected type of vulnerability. For the multiclass vulnerability classification task, we used our curated dataset VulF and MVD.
We also tested our classifier's ability to separate vulnerable and non-vulnerable classes as a binary classification task. We tested the performance of binary classification using three datasets; namely, FFMpeg+Qemu \cite{zhou2019devign}, ReVEAL \cite{chakraborty2021deep}, and D2A \cite{zheng2021d2a}.

\paragraph{\textbf{Discussion}}
In order to analyze the performance of our multitask vulnerability description model, our model was tasked with classifying 40 categories of vulnerability \color{black} from VulF dataset where each category is associated with a CWE number and a description. For multiclass vulnerability classification, the goal of the classifier initially is to detect whether vulnerability exists in the code, and if a vulnerability exists, provide the CWE number as depicted in Figure \ref{fig:code}. This experiment tests our classifier's ability to detect and classify vulnerable code samples \color{black}. Table \ref{fig:classification} shows a comparison of the classification performance of various models against ours when tested with different datasets. Table \ref{tab:multiclassvuln} shows that our VulF dataset is mildly imbalanced. For example, the number of vulnerable codes for CWE-676, CWE-362, and CWE-662 is deficient, less than 300. On the other hand, there are 4713 code samples for CWE-119 and 1380 for CWE-704. We prevented training a biased model by using the Focal Loss function during the model's training.

\begin{table}[t]
\centering
\caption{\color{black}
 Comparison of our proposed RoBERTa-PFGCN 
 vulnerability detection model against top recent models, including $\mu$VulDeePecker, BiLSTM, TextCNN, RoBERTA, CodeBERT, Devign, and VELVET, with a focus on their respective datasets.\color{black}}
\label{fig:classification}
\begin{tabular}{p{0.1\textwidth}
                 p{0.11\textwidth}
                 p{0.03\textwidth}
                 p{0.03\textwidth}
                 p{0.03\textwidth}
                 p{0.03\textwidth}}
\toprule
\textbf{Data}                           & \textbf{Model}                              & \textbf{Acc.}            & \textbf{Prec}      & \textbf{Recall}         & \textbf{F1}                  \\ \midrule

\multirow{4}{*}{VulF} & $\mu$VulDeePecker     & 78.35          & 78.94          & 78.30          & 77.10 \\

& Devign            & 84.55          & 83.94          & 83.15          & 84.30          \\

& VELVET            & 84.45          & 85.12          & 85.20          & 84.46          \\

& \textbf{RoBERTa-PFGCN} & \textbf{96.24} & \textbf{96.18} & \textbf{95.15} & \textbf{95.85}         \\ \bottomrule


\multirow{2}{*}{MVD }  & $\mu$VulDeePecker \cite{zou2019mu}   & -          & -          & -          & 94.22          \\
                               & \textbf{RoBERTa-PFGCN(Ours)}         & 98.23 & 98.28  & 98.23  & \textbf{98.01}  \\

\toprule

\multirow{6}{*}{FFMpeg + Qemu} & BiLSTM \cite{lu2021codexglue}      & 59.37          & -              & -              & -              \\
                               & TextCNN \cite{lu2021codexglue}     & 60.69          & -              & -              & -              \\
                               & RoBERTa \cite{lu2021codexglue}     & 61.05          & -              & -              & -              \\
                               & CodeBERT \cite{lu2021codexglue}    & 62.08          & -              & -              & -              \\
                               & Devign \cite{chakraborty2021deep}  & 58.57          & 53.60          & 62.73          & 57.18          \\
                               & \textbf{RoBERTa-PFGCN(Ours)}         & \textbf{63.29} & \textbf{63.08} & \textbf{62.97} & \textbf{62.99} \\
\toprule

\multirow{2}{*}{D2A}           & VELVET \cite{ding2021velvet}        & 59.3           & \textbf{70.5}  & 50.4            & 58.8          \\
                               & \textbf{RoBERTA-PFGCN(Ours)}         & \textbf{61.2}  & 61.97          & \textbf{62.07}  & \textbf{61.21}\\ \bottomrule
\end{tabular}
\end{table}

\color{black}
We tested VulF dataset with our proposed RoBERTa-PFGCN and two other publicly available models, Devign and VELVET. Moreover, we also implemented $\mu$VulDeePecker \cite{zou2019mu} ourselves based on their proposed architecture since their model is not publicly available. Our experimental results show that our model achieves the highest accuracy compared to Devign, VELVET, and $\mu$VulDeePecker. Table \ref{fig:classification} shows that our model improved accuracy, precision, recall, and F1 score by 11.79\%,  11.06\%,  9.98\%, and  11.39\%, respectively, which is at least 11\% higher than other models.
\color{black}


Our work was also compared with the results from $\mu$VulDeePecker \cite{zou2019mu}, which evaluated 40 classes. Table \ref{fig:classification} shows that our model outperforms $\mu$VulDeePecker \color{black} on their MVD dataset by almost 3.80\% \color{black} on F1 score. \color{black} We provide an F1 score comparison for all 40 classes by our model against $\mu$VulDeePecker with the MVD dataset in Figure \ref{fig:40_class}. This bar chart shows a comparative analysis of the F1 score with our model vs. $\mu$VulDeePecker, and we can clearly see our model consistently generates higher F1 scores for all the 40 CWE classes.
\color{black}

Table \ref{fig:classification} demonstrates our model's performance with two other datasets. These results also illustrate that baseline models like BiLSTM and TextCNN significantly underperform compared to the pre-trained programming language (PL) models like CodeBERT and RoBERTa, as well as our proposed model Multitask RoBERTa-PFGCN, with the FFMpeg+Qemu dataset. Compared to non-PL-based models, our model shows an improvement of 3.92\% and 2.60\% over BiLSTM and TextCNN and 2.24\%, 1.21\%, and 4.72\% compared to  RoBERTa, CodeBERT, and Devign, respectively.

For the D2A dataset, \color{black} we compared our work with VELVET \cite{ding2021velvet}, and see \color{black} an improvement in classification Accuracy by 1.9\%, Recall by 11.67\%, and F1 score by 2.41\%.


\textbf{RQ2: Can our classifier learn vulnerability in a biased setting?}

\color{black}
An important limitation of real-world source code data is its imbalanced nature. Since real-world projects have very few vulnerable but very many non-vulnerable programs, vulnerability models suffer from data imbalance \cite{chakraborty2021deep}. The imbalance scenario renders the model more biased towards the non-vulnerable class rendering a higher accuracy with lower precision and recall scores. Of the datasets we have discussed thus far, ReVEAL is highly imbalanced; hence have used it to test against biased settings. \color{black}
A sound vulnerability system should not have a poor F1, Precision, and Recall score despite a potentially high Accuracy, as this would prove bias towards the majority class. We investigated the usefulness of the Focal Loss \cite{lin2017focal} function and how they prevent biases in the model. For this experiment, we fixed the weight hyper-parameters $\alpha = 0.1$ and $\gamma = 2$ \color{black}for ReVEAL dataset with our RoBERTa-PFGCN model and SVG as input\color{black}. Since Focal Loss uses a cross-entropy loss function underneath its implementation, we set the learning rate to $1e-5$ with a batch size of 256.

\paragraph{\textbf{Discussion}} We used the ReVEAL dataset for this experiment, which has a 1 to 9 ratio of vulnerable to non-vulnerable code. Table \ref{fig:focal} shows that the works proposed by Russell et al.\cite{russell2018automated}, VulDeepecker \cite{li2018vuldeepecker} SySeVR \cite{li2021sysevr} and Devign \cite{zhou2019devign} achieve high accuracy, but their Precision, Recall, and F1 scores drop significantly.\color{black} To overcome this issue, Focal Loss is used to add more weight to the loss if the model incorrectly predicts the minority class. Thus, we set the value of weight  $\alpha$ by taking an inverse ratio of vulnerable to non-vulnerable example codes during training and we set the value of $\gamma$ to 2 by experimental analysis. However, $\gamma$ is used as an exponent in Equation \ref{eqn:8}, so when we run RoBERTa-GCN w/ WL, $\gamma$ is set to 0 in order to ignore its effect. Observing the last two rows in Table \ref{fig:focal}, we find that, initially, we tested our model using Weighted Loss (WL) only, and later we tested with Focal Loss, which is a combination of weighted loss $\alpha$ with the hyperparameter $\gamma$. \color{black}

In both cases, we observe that our model has surpassed previous models in terms of Precision, Recall, and F1 score, indicating lower false positive and false negative rates with WL and with FL. However, these numbers improved slightly when compared with performance between weighted loss and Focal Loss, with the exception of the precision metric. From the results, it is observed that with Focal Loss, we achieved an improvement of 11.18\% in Precision, 1.06\%  in Recall, and 0.61\% on our F1 score compared to the weighted loss. Nevertheless, using either weighted loss or Focal Loss, the accuracy of our model drops by almost 1\% compared to \cite{russell2018automated}, indicating that these previous models are highly biased with an assumption of non-vulnerability. Compared to the next best model \cite{zhou2019devign}, our model with Focal Loss shows an improvement of 22.91\% in Precision, 23.53\% in Recall, and 18.04\% on F1 score. \color{black}Thus, Focal Loss improves precision, recall and F1 on imbalanced data while not causing a negating impact on a balanced dataset only by adjusting the parameters $\alpha$ and $\gamma$ \color{black}

\begin{table}[t]
\centering
\caption{Vulnerability Classification Using Focal Loss. This table shows the effectiveness of our model using the Focal Loss function. Since only ReVEAL is imbalanced, this dataset is used for result comparison. Results are taken from \cite{chakraborty2021deep}.}
\label{fig:focal}
\begin{tabular} {p{0.11\textwidth}
                 p{0.12\textwidth}
                 p{0.03\textwidth}
                 p{0.03\textwidth}
                 p{0.03\textwidth}
                 p{0.03\textwidth}}
\toprule
\textbf{Model}                      & \textbf{Input}                          & \textbf{Acc.}          & \textbf{Prec.}         & \textbf{Recall}        & \textbf{F1}    \\ \midrule
Russell et al.\cite{russell2018automated}             & Token                          & \textbf{90.98}         & 24.63         & 10.91         & 15.24 \\
VulDeePecker \cite{li2018vuldeepecker}               & \multirow{2}{*}{Slice + Token} & 89.05         & 17.68         & 13.87         & 15.7  \\
SySeVR \cite{li2021sysevr}                    &                                & 84.22         & 24.46         & 40.11         & 30.25 \\
Devign \cite{zhou2019devign}                     & CPG                            & 88.41         & 34.61         & 26.67         & 29.87 \\
\textbf{RoBERTa-PFGCN w/ Weighted Loss} & Semantic Vulnerability Graph                 & 88.07         &46.34          & 49.14         & 47.31 \\
\textbf{RoBERTa-PFGCN w/ Focal Loss} & Semantic Vulnerability Graph       & 89.88 & \textbf{57.52} & \textbf{50.20} & \textbf{47.91}     \\ \bottomrule

\end{tabular}
\end{table}


\textbf{RQ3: Is our classifier generalized enough to detect vulnerabilities in \color{black} N-day and zero-day program samples\color{black}?}

We evaluated our classifier's performance based on its ability to accurately predict vulnerability with \color{black}273 N-day \color{black} real-world sample programs. These sample programs are never used during training. \color{black} We also used 4 zero-day examples in order to evaluate our classifier on predicting zero-day vulnerabilities as well. \color{black}The classifier predicts the vulnerability \color{black}class\color{black} if the vulnerability exists in the code and predicts non-vulnerable when the vulnerability does not exist. Out of these \color{black}273 N-day and 4, zero-day code samples, some vulnerability classes exist that are not part of our VulF dataset from table \ref{tab:multiclassvuln}\color{black}. For example, the VulF dataset has no code samples for CWE-787 for training. But a few samples from our VulF dataset have code samples for CWE-787. So when we evaluated our classifier \color{black}on 273, N-day and 4, zero-day code samples,\color{black} our model could not predict the vulnerability classification of that particular case. Table \ref{tab:0N-Day} provides a more in-depth analysis of our work with the recent models like $\mu$VulDeepecker \cite{zou2019mu}, Devign \cite{zhou2019devign} and VELVET \cite{ding2021velvet}. We can see that our model was able to detect most N-day and zero-day vulnerabilities compared to previous works.\color{black}

\paragraph{\textbf{Discussion}}
In this experiment, we observed that our model achieves an accuracy of 93.00\% when trained with our VulF dataset and tested against N-day sample programs of 273 examples. Out of the three datasets that we have used for different experiments, the VulF dataset shows the best performance when we trained our model with SVG. Out of the 273, \color{black} N-day code samples \color{black} our model can successfully predict 255 as vulnerable, achieving an accuracy of 93\%.  \color{black}Moreover, for zero-day analysis, we tested the same model for 4, zero-day examples, and out of those, our model was able to predict all examples correctly.  Furthermore, we provide four case studies, 3, N-day and 1, zero-day program samples, with a critical analysis and the reasoning behind our classifier's outcome. \color{black}


\begin{figure}[b]
        \centering
        
        \includegraphics[width=1\columnwidth]{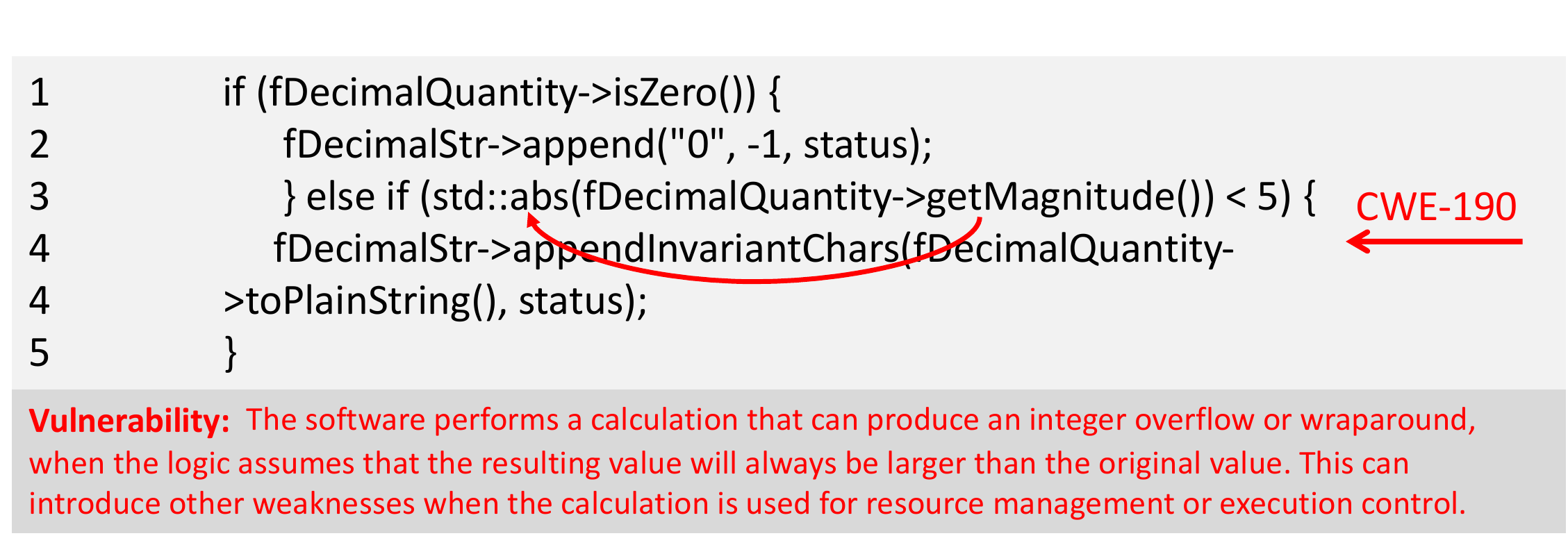}
        
    \caption{An example code for CWE-190, which our classifier predicted accurately. The red edge shows a Poacher Flow edge that captures the Data Processing of the code. Hence, our classifier was able to detect the vulnerability with a description.}
    \label{fig:case_1}
\end{figure}

\begin{table}[t]
\centering
\caption{\color{black}N-day and zero-day comparison of our work with previous works. \color{black}}
\label{tab:0N-Day}
\begin{tabular}{p{0.10\textwidth}|
                 p{0.02\textwidth}|
                 p{0.01\textwidth}|
                 p{0.02\textwidth}|
                 p{0.01\textwidth}}
\toprule
\textbf{Model}                         & \multicolumn{1}{l|}{\textbf{N-day}} & \multicolumn{1}{l|}{\textbf{\begin{tabular}[c]{@{}l@{}}Correctly \\ Predicted\end{tabular}}} & \multicolumn{1}{l|}{\textbf{zero-day}} & \multicolumn{1}{l}{\textbf{\begin{tabular}[c]{@{}l@{}}Correctly\\  Predicted\end{tabular}}} \\ \hline
Devign                          & \multirow{3}{*}{273}                & 202                                                                                           & \multirow{3}{*}{4}                 & 1                                                                                          \\ \cline{3-3} \cline{5-5} 

VELVET                                 &                                     & 208                                                                                           &                                     & 1                                                                                          \\ \cline{1-1} \cline{3-3} \cline{5-5} 
\multicolumn{1}{l|}{\textbf{RoBERTa-PFGCN}} &                                     & \textbf{255}                                                                                           &                                     & \textbf{4}                                                                                          \\ \bottomrule
\end{tabular}
\end{table}


\begin{figure}[t]
        \centering
        
        \includegraphics[width=1\columnwidth]{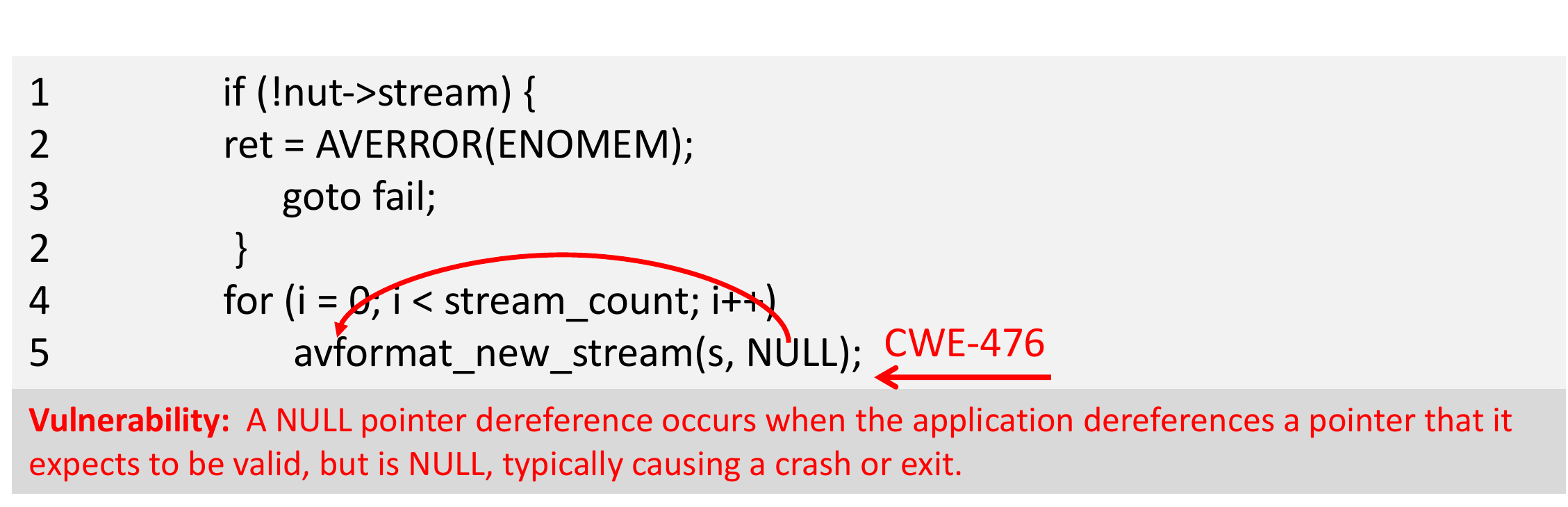}
        
    \caption{A sample code for CWE-476, which our classifier accurately predicted. The red edge shows a Poacher Flow edge that captures the Access Control of the code. Thus, our classifier was able to detect the vulnerability with a description.}
    \label{fig:case_2}
\end{figure}


\paragraph{\textbf{Case Study 1}} \color{black}We collected this N-day sample program from International Components for Unicode (ICU) repository. In this example, the program in Figure \ref{fig:case_1} attempts to get the value of $fDecimalQuantity$ without checking the possible integer limit. As a result, a potential buffer overflow could potentially crash the program when the $abs$ function is called. The red edge in Figure \ref{fig:case_1} shows the Data Processing Edge, which hints to the classifier that a vulnerability may exist. Thus, the classifier emphasized the information provided by this edge, detected the code as vulnerable, and classified the vulnerability as CWE-190\color{black}.


\paragraph{\textbf{Case Study 2}}
We collected this N-day sample program from the FFMpeg repository from GitHub. Here the method $avformat\_new\_stream$ is called without checking the possible value of the output. The output can potentially trigger a null pointer dereference error causing the application to crash. When this code goes through our classifier, the classifier observes the PF edge, observed in red in Figure \ref{fig:case_2} (Data Processing edge), and classifies this code as vulnerable (CWE-476).

\begin{figure}[b]
        \centering
        
        \includegraphics[width=1\columnwidth]{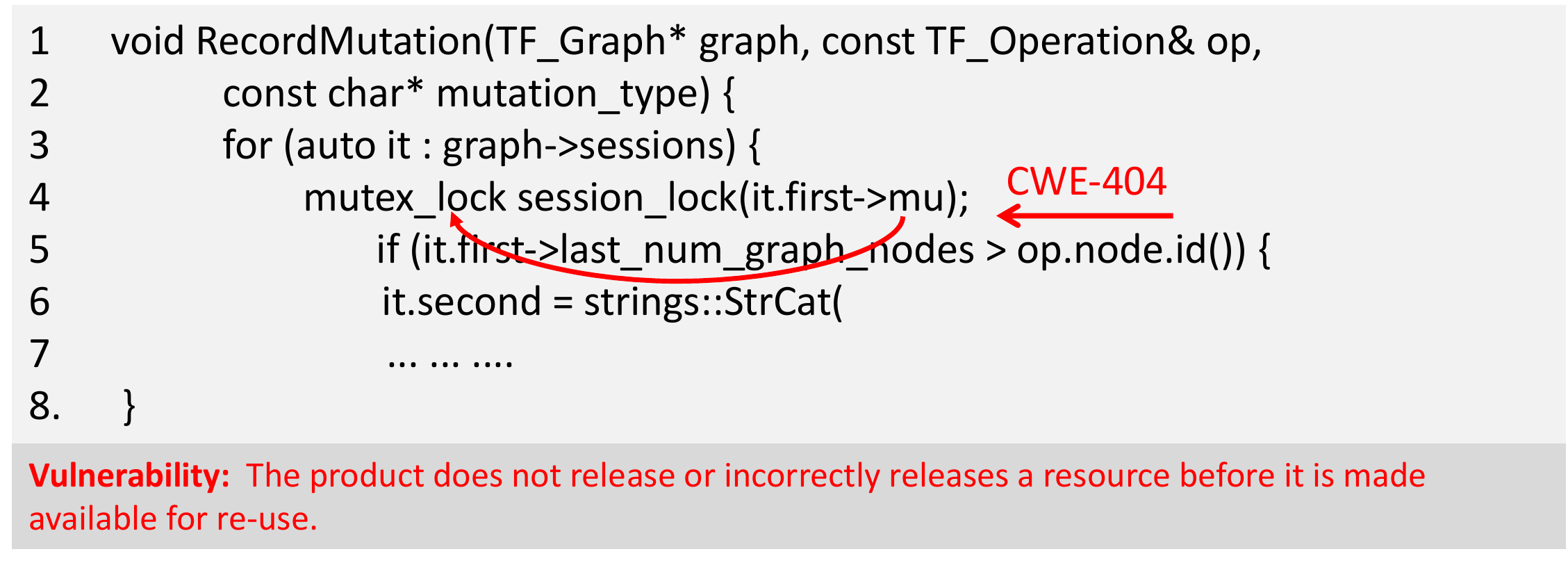}
        
    \caption{A sample code for CWE-404, which our classifier accurately predicted. The red edge shows a Poacher Flow edge that captures the Resource Management of the code. Thus, our classifier was able to detect the vulnerability with a description.}
    \label{fig:case_3}
\end{figure}

\paragraph{\textbf{Case Study 3}} 
This zero-day vulnerable example is also part of the TensorFlow C repository. Figure \ref{fig:case_3} shows a fraction of a large function used during evaluation. In line 4, the code tries to create a mutex lock by invoking $mutex\_lock\_session$. However, till the end of the function, there was no call to unlock the resource $it.first->mu$. As a result, this resource is locked indefinitely, causing a resource management issue. Since a Resource Management edge exists for this situation, our classifier detects the vulnerability and classifies it as CWE-404, which eventually creates a deadlock situation.

\begin{figure}[t]
        \centering
        
        \includegraphics[width=1\columnwidth]{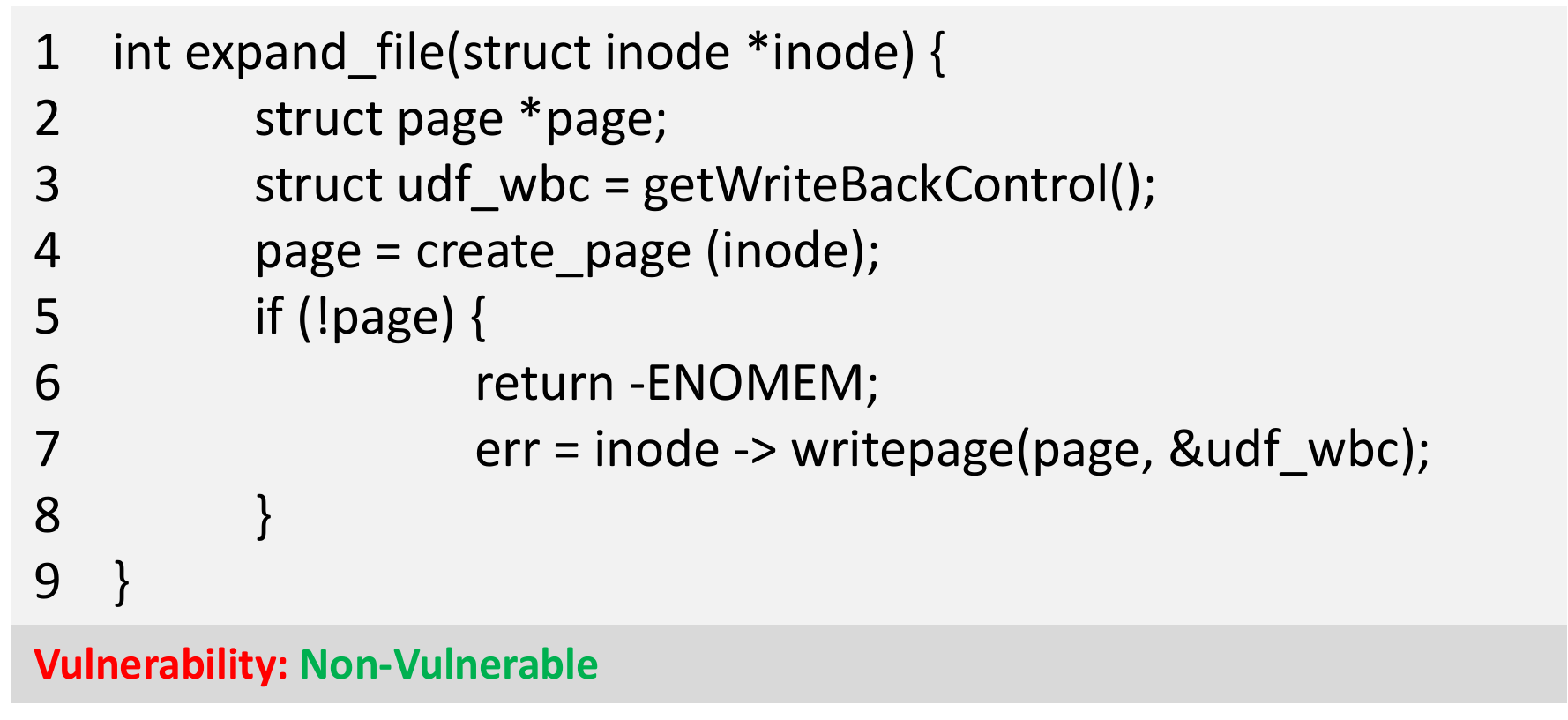}
        
    \caption{An example code for CWE-476 that our classifier could not predict accurately. No poacher edges exist for this code. Hence our model predicted it as Non-Vulnerable}
    \label{fig:case_4}
\end{figure}

\paragraph{\textbf{Case Study 4}}  \color{black}We collected this N-day, sample program from Linux repository. \color{black} From GitHub commit messages, we learned that this vulnerability occurs when the function attempts to write data expanded into a page but fails to set up the \textit{inode}, which serves as a unique identifier for information on a specific filesystem. Consequently, a null pointer dereferencing error could occur during writeback if the inode is not created. Figure \ref{fig:case_4} shows that the statement \textit{writepage} attempts to write a page using the parameters \textit{page and udf\_wbc}. Nevertheless, a checking statement (line 5) already exists before assigning a value in line 7. Our classifier identifies the code snippet as non-vulnerable as it is aware that the parameters are not null. However, a different API call (\textit{filemap\_fdatawrite}) generates this vulnerability (CWE-476) as a result of an inappropriate request for the overall task. Since no logical Poacher edge could be identified for this vulnerability and the other edge do not contribute to detecting vulnerabilities from the implementation of API $writepage$, our model incorrectly classifies this code as non-vulnerable.

%

\begin{table}[b]
\centering
\caption{Comparing the execution times of Generating SVG and AST Graph}
\label{tab:ablation}
\begin{tabular}{p{0.10\textwidth}
                 p{0.13\textwidth}
                 p{0.06\textwidth}}

\toprule
\textbf{Training Data}              & \textbf{Graph}   &  \textbf{Execution Time}     \\ \midrule

\multirow{2}{*}{VulF}        & SVG           & 2m 32s            \\
                                    & AST               & 18m        \\  \midrule

\multirow{2}{*}{MVD}                & SVG           & 28m 53s             \\
                                    & AST             & 3m 18         \\ \midrule

\multirow{2}{*}{FFMpeg+Qemu}        & SVG          & 38s            \\
                                    &AST               & 7m 20s        \\  \midrule

\multirow{2}{*}{D2A}                & SVG          & 11s             \\
                                    & AST              & 2m 52s         \\ \midrule

\multirow{2}{*}{ReVEAL}             & SVG           & 20s            \\
                                    & AST           & 4m 15s         \\ 

                                    \bottomrule
\end{tabular}
\end{table}

\color{black}
\subsection{Time and Memory Analysis}
\color{black}
In this section,  we analyze model complexity in terms of memory consumption and processing time. We assessed our classifier using AST and SVG. We generated an AST using a tool called TreeSitter \cite{tree}. Table \ref{tab:ablation} demonstrates that AST generation incurs significant time and memory overhead. It takes approximately 18 minutes to construct the AST of 141285 functions but only 2 minutes and 32 seconds to generate our SVG for the same VulF dataset. We observe a similar case for the ReVEAL and the D2A datasets. In the ReVEAL dataset, the AST input pre-processing time is 13 times higher than the creation time of our SVG. In the D2A dataset, the AST pre-processing time is 16 times higher. \color{black} Finally, for the MVD dataset, the AST pre-processing time is 11 times higher. \color{black}

Furthermore, we discovered high memory overhead issues during the creation of the ASTs.  For their generation, a process must traverse the tree recursively in order to build it. As a result, the internal stack expands exponentially as the size of the program increases. In comparison, our SVG doesn't rely on a recursive program for its generation. We perform our analysis in Python and Pytorch. Python's standard stack size is 1,000. However, the stack size had to be increased to 3,000 to build ASTs for all the functions presented in the different datasets we used. A larger stack size causes at least three times more memory consumption when the function grows too large. 

%
\begin{table}[t]
\centering
\caption{\color{black} In-depth ablation study for each component of our proposed PF Edges\color{black}}
\label{tab:ablation_2}
\begin{tabular}{p{0.08\textwidth}
                 p{0.14\textwidth}
                 p{0.03\textwidth}
                 p{0.02\textwidth}
                 p{0.02\textwidth}
                 p{0.02\textwidth}}
\toprule
\textbf{Dataset}                                 & \multicolumn{1}{l}{\textbf{Model}} & \multicolumn{1}{c}{\textbf{Acc.}} & \multicolumn{1}{c}{\textbf{Prec.}} & \multicolumn{1}{c}{\textbf{Recall}} & \multicolumn{1}{c}{\textbf{F1}} \\ \midrule

\multirow{6}{*}{\begin{tabular}[c]{@{}l@{}}VulF\end{tabular}} & GCN                           & 87.50                                      & 87.36                                       &                  87.94                  &       87.23       \\

& GCN w/ AC    & 91.23                                      & 91.75                                       &                  91.56                   &       91.23       \\

& GCN w/ DP     & 90.40                                      & 89.65                                       &                  90.95                   &       90.50       \\

& GCN w/ RM    & 89.54                                      & 90.28                                       &                  90.61                   &       90.83       \\

&  \textbf{RoBERTa-PFGCN}  & \textbf{96.24}                                    &               \textbf{96.18}&     \textbf{95.15}&                       \textbf{95.85} \\
\hline


\multirow{6}{*}{\begin{tabular}[c]{@{}l@{}}MVD\end{tabular}} & GCN                           & 86.10                                      & 85.75                                       &                  85.24                  &       86.50       \\
& GCN w/ AC    & 92.16                                      & 91.90                                       &                  91.10                   &       91.70       \\              
& GCN w/ DP     & 91.70                                      & 90.15                                       &                  91.57                   &       90.30       \\                    
& GCN w/ RM    & 91.46                                      & 92.40                                       &                  91.43                   &       91.33       \\                    
&  \textbf{RoBERTa-PFGCN}&\textbf{98.23}&\textbf{98.28}&\textbf{98.23}&\textbf{98.01}\\
\hline



\multirow{6}{*}{\begin{tabular}[c]{@{}l@{}}FFMpeg +\\Qemu\end{tabular}} & GCN                           & 56.34                                      & 57.47                                       &                  57.28                  &       56.57       \\
& GCN w/ AC    & 60.94                                      & 60.12                                       &                  60.35                   &       60.48       \\              
& GCN w/ DP     & 61.50                                      & 60.51                                       &                  61.23                   &       60.80       \\                    
& GCN w/ RM    & 61.90                                      & 61.08                                       &                  61.40                   &       61.74       \\                    
& \textbf{RoBERTa-PFGCN}                                     &\textbf{63.29}&\textbf{63.08}&\textbf{62.97}&\textbf{62.99}\\
\hline



\multirow{6}{*}{\begin{tabular}[c]{@{}l@{}}D2A\end{tabular}} & GCN                           & 57.95                                      & 58.80                                       &                  57.69                  &       58.46       \\
& GCN w/ AC    & 61.29                                      & 62.50                                       &                  61.58                   &       62.20       \\              
& GCN w/ DP     & 61.65                                      & 60.10                                       &                  61.95                   &       60.28       \\                    
& GCN w/ RM    & 60.30                                      & 59.28                                       &                  59.10                   &       59.25       \\                    
& \textbf{RoBERTa-PFGCN}&\textbf{61.20}&\textbf{61.97}&\textbf{62.07}&\textbf{61.21}\\


\bottomrule

\end{tabular}
\end{table}

\color{black}
\subsection{Ablation Studies}


We evaluated  our model's performance with and without the influence of Poacher Flow (PF) edges. As observed in Table \ref{tab:ablation_2}, we trained and evaluated the performance of our classifier on four datasets to ensure that the observed performance improvement was not a coincidental occurrence. The experimental hyperparameters were similar across the experiments we performed with the exception of the hyperparameters tailored for \textbf{RQ2}. Initially, we trained and evaluated two models, RoBERTa-PFGCN and RoBERTa-GCN, with and without PF edges, respectively. Afterwards, we trained new versions of the RoBERTa-GCN model by adding each component of our PF edges, such as the \textit{Data Processing (DP) Edge}, \textit{Access Control (AC) Edge}, and \textit{Resource Management (RM Edge} to understand the contribution that each PF component had on the model's performance. As seen in Table \ref{tab:ablation_2}, our proposed RoBERTa-PFGCN vulnerability detection model achieves better performance when trained with PF edges resulting in an 8.74\% improvement in accuracy and 8.62\% in F1 score when trained and tested using the VulF dataset. 


\color{black}

\section{Conclusion and Future Work}
\label{6_conclusion}


This paper employed a unique set of edges, including novel Poacher Flow edges to generate richer vulnerability detection and description features. With Poacher Flow edges, our classifier can detect vulnerability \color{black}that may arise due to the dynamic behavior of a program during execution and assignments. We propose a set of algorithms to generate PF edges from source code.\color{black} We used a classification model for detecting and classifying source code vulnerabilities based on Multitask RoBERTa-GCN with a Focal Loss and their corresponding classification. We utilized Focal Loss to rectify our model's bias toward the majority class, decreasing our model's false positives and false negatives and resulting in state-of-the-art source code vulnerability detection for real-world projects.\color{black} We also provided an in-depth ablation study for evaluating the performance impact that each component of our PF edges has on our model. In addition, we performed a time and memory analysis for the generation of ASTs and our SVG. \color{black}Finally, we introduced the VulF dataset which provides software developers with vulnerability detection and CWE description, helping them resolve source code vulnerability issues. Our future work will focus on reasoning and counterfactual explanations for code vulnerability localization and corrections.


\color{black}

\section*{Acknowledgements}
This research project and the preparation of this publication were funded in part by the National Science Foundation under Grant No. 2230086.

\bibliographystyle{unsrt}

\bibliography{main}





\end{document}